\documentclass[12pt]{article}

\usepackage{graphics}
\usepackage{amsthm,amssymb,mathrsfs,setspace,amsmath}
\numberwithin{equation}{section}
\usepackage{graphicx}
\usepackage{booktabs}
\usepackage{tikz}
\usepackage{multirow}
\usepackage{setspace}
\usepackage{subcaption}
\usepackage[english]{babel}

\title{ Reliability Analysis of Load-sharing Systems using a Flexible Model with Piecewise Linear Functions}
\author{Shilpi Biswas \thanks{Indian
Institute of Technology Guwahati, Assam 781039, India; Email:
shilpi.biswas@iitg.ac.in}, 
Ayon Ganguly \thanks{Indian
Institute of Technology Guwahati, Assam 781039, India; Email:
aganguly@iitg.ac.in}, 
and
Debanjan Mitra \thanks{Indian Institute of Management Udaipur, Rajasthan 313001,
India; Email: debanjan.mitra@iimu.ac.in}}
\date{}

\allowdisplaybreaks

\begin{document}

\maketitle

\begin{abstract}
Aiming for accurate estimation of system reliability of load-sharing systems, a flexible
model for such systems is constructed by approximating the cumulative hazard functions of
component lifetimes using piecewise linear functions. The advantages of the resulting
model are that it is data-driven and it does not use prohibitive assumptions on the
underlying component lifetimes. Due to its flexible nature, the model is capable of
providing a good fit to data obtained from load-sharing systems in general, thus resulting
in an accurate estimation of important reliability characteristics. Estimates of
reliability at a mission time, quantile function, mean time to failure, and mean residual
time for load-sharing systems are developed under the proposed model involving piecewise
linear functions. Maximum likelihood estimation and construction of confidence intervals
for the proposed model are discussed in detail. The performance of the proposed model is
observed to be quite satisfactory through a detailed Monte Carlo simulation study.
Analysis of a load-sharing data pertaining to the lives of a two-motor load-sharing system
is provided as an illustrative example. In summary, this article presents a comprehensive
discussion on a flexible model that can be used for load-sharing systems under minimal
assumptions.      
\end{abstract}

\noindent\textbf{Keywords:} Load-sharing systems; Cumulative hazard function; Baseline hazard; Piecewise linear approximation; Maximum likelihood estimation; Fisher information; Bootstrap; Confidence interval; Quantile function; Mean time to failure; Reliability at a mission time; Mean residual time.

\section{\sc Introduction}\label{sec:intro}
\subsection{\sc Background}
Dynamic models are suitable for reliability systems where failure or degradation of one or
more components affects the performance of the surviving or operating components. Load-sharing systems are appropriate examples where such models can be used. The total load on a load-sharing system is shared between its components; when a component fails within the system, the total load gets redistributed over the remaining operating components. As a result of a higher stress due to this extra load, the failure rates of the operating components increase. 

Common examples of load-sharing systems are those where components are connected in
parallel, such as central processing units (CPUs) of multi-processor computers, cables of
a suspension bridge, valves or pumps in hydraulic systems, electrical generator systems
etc. Load-sharing systems are found in other spheres as well, such as the kidney system in
humans. When one of the kidneys fails or deteriorates, the other kidney experiences
elevated stress and has an increased chance of failure. 

The load-share rule among the operating components depends on the physical characteristics
of the system involved. In an equal load-share rule, the extra load caused by the failed
components is shared equally by the operating components. On the other hand, a local
load-share rule implies that the extra load is shared by the neighboring components of the
failed ones. A monotone load-sharing rule more generally assumes that the load on the
operating components is non-decreasing with respect to the failure of other components in
the system \cite{Kvam2008}. 

\subsection{\sc Literature review}

One of the early major contributions to the literature on load-sharing systems was by
Daniels~\cite{Daniels1945}, describing the increasing stress on yarn fibres with
successive breakings of individual fibres within a bundle. In the same context of the
textile industry, the early-period literature saw developments by
Coleman~\cite{coleman1957a, coleman1957b}, Rosen~\cite{Rosen1964}, and Harlow and
Phoenix~\cite{Harlow1978, Harlow1982}, among others. In general, the topic attracted the
attentions of several researchers, and significant theoretical contributions were made, for example, by Birnbaum and Saunders~\cite{Birnbaum1958}, Freund~\cite{F1961}, Ross~\cite{Ross1984}, Schechner~\cite{Schechner1984}, Lee et al.~\cite{LeeDurham1995}, Hollander and Pena~\cite{Hollander1995}, and Lynch~\cite{Lynch1999}. 

While most studies on load-sharing systems in the early-period were based on a known
load-share rule, Kim and Kvam~\cite{KimKvam2004} presented a statistical methodology for
multicomponent load-sharing systems with an unknown load-share rule. In fact, the work of
Kim and Kvam~\cite{KimKvam2004} was also important for another reason: they used the
hypothetical latent variable approach for modelling the component lifetimes. The latent
variable approach was later adapted by Park~\cite{Park2010, Park2013} for developing an
inferential framework for load-sharing systems assuming the component lifetimes to be
exponential, Weibull, and lognormally distributed random variables.

The use of parametric models has a long history in the literature on load-sharing models.
Exponential distribution has been extensively used for modelling lifetimes of components
of load-sharing systems \cite{SL1991, LCW1993, MKA2012}. However, the property of a
constant hazard rate of the exponential distribution is not practical for most
applications. The tampered failure rate model for load-sharing systems, proposed by
Suprasad et al.~\cite{SKH2008}, was thus developed to accommodate a wide range of
failure-time distributions for the components. In this connection, the use of accelerated
life testing models for load-sharing systems may be mentioned; see Mettas and
Vassiliou~\cite{Mettas2004}, Amari and Bergman~\cite{Bergman2008}, and Kong and
Ye~\cite{Kong2016}. A family of parametric distributions was used for modelling the lives of
two-component load-sharing systems by Deshpande et al.~\cite{DDN2010}. Asha et
al.~\cite{ARR2018} used a frailty-based model to this effect. A recent contribution in
this direction is by Franco et al.~\cite{FVK2020} who used generalized Freund's bivariate
exponential model for two-component load-sharing systems. See also the references cited in
these articles. 


Recently, several authors have explored diverse areas concerning load-sharing systems. The damage accumulation of load-sharing systems was modelled by M\"{u}ller and Meyer~\cite{MM2022}. Luo et al.\cite{LSX2022} developed a model for correlated lifetimes in dynamic environments incorporating the load-sharing criterion. 
Brown et al.~\cite{BLMR2022} explored a spatial model for load-sharing where the extra
load due to failure of a component is shared more by the operating components that are in
close proximity of the failed component than those that are distant. Nezakati and
Ramzakh~\cite{NR2020}, and Zhao et al.~\cite{ZLL2018} connected degradation of components
to load-sharing phenomena. In an interesting development, Che et al.~\cite{CZLG2022}
considered man-machine units (MMUs) as units of analysis where load-sharing was possible
due to machine issues as well as human issues. They studied the load-sharing of the MMUs, attempting to capture the complex dependence between machines and their operators. A general model, called the load-strength model, was studied by Zhang et al.~\cite{ZZM2020}.     It is to be noted that most of the studies on load-sharing systems have used parametric models for analysis so far, thus heavily relying on the modelling assumptions for suitability of their analyses.

\subsection{\sc Aim and Motivation}
Our aim in this paper is to develop an appropriate estimate for the system reliability or
reliability at mission time (RMT) of load-sharing systems. The aim, also, is to accurately
estimate quantile function of the underlying system lifetime distribution, mean time to
failure (MTTF), and mean residual time (MRT) of load-sharing systems. These quantities are
important to fully understand the characteristics of a load-sharing system; also, they are
of practical importance for making various strategies and plans. 

Naturally, the quality of estimation of RMT, quantile function, MTTF, and MRT of a
load-sharing system depends on the suitability of the model that is fitted to the
lifetimes of its components capturing the load-share rule. To this effect, we develop a
model for the component lifetimes involving piecewise linear approximations (PLAs) of the
cumulative hazard functions, capturing the unknown load-share rule at each of the
successive stages of component failures. The model is data-driven, and does not require
prohibitive parametric assumptions for component lifetime distributions. Due to this
flexibility, the PLA-based model is capable of providing a good fit to load-sharing data.
An example, elaborated in a later section, is as follows. 

Data pertaining to a load-sharing system where each system was a parallel combination of
two motors were analysed by Asha et al.~\cite{ARR2018} and Franco et al.~\cite{FVK2020}.
Asha et al.~\cite{ARR2018} assumed Weibull distributions for the component lifetimes,
although data for one of the two component motors showed clear empirical evidence that the
assumption was not satisfied. A generalized bivariate Freund distribution was assumed for
the component lifetimes by Franco et al.~\cite{FVK2020}. To this data, we have fitted our
proposed PLA-based model, and have observed according to the Akaike's information
criterion (AIC) for model selection, the PLA-based model is a much better fit compared to
the Weibull model of Asha et al.~\cite{ARR2018} and generalized bivariate Freund model of
Franco et al.~\cite{FVK2020}. The immediate and obvious result of this is a much more
accurate estimation of the RMT, quantile function, MTTF, and MRT of the system lifetimes.
The details of this analysis are given in a later section. 

The main contributions of this paper are as follows: 
\begin{itemize}
	\item We develop a flexible, data-driven model based on PLA for modelling component lifetimes of a load-sharing system. The model does not require prohibitive parametric assumptions on the underlying component lifetimes.
	\item We develop inference for the proposed PLA-based model based on data from multi-component load-sharing systems.
	\item Under the proposed PLA-based model, we develop methods to accurately estimate
      important reliability characteristics such as system reliability or RMT, quantile function, MTTF, and MRT of load-sharing systems. 
\end{itemize}

The rest of this article is structured as follows. In Section \ref{sec:Model}, the proposed PLA-based model for load-sharing systems is presented. Section \ref{sec:Inf} contains likelihood inference for the model based on data from multi-component load-sharing systems, including relevant details of derivation of MLEs, construction of confidence intervals, and a general guidance on selection of cut-points for the piecewise linear functions. Estimation of system reliability, quantile function, MTTF, and MRT of load-sharing systems in this setting are given in Section \ref{sec:RC}. Based on component lifetime data from a two-component load-sharing system, an illustrative example of application of the PLA-based model and estimation of various important reliability characteristics are presented in Section~\ref{sec:dataana}. In Section \ref{sec:sim}, results of a detailed Monte Carlo simulation experiment investigating the efficacy and robustness of the PLA-based model are presented. Finally, the paper is concluded with some remarks in Section \ref{sec:con}.    

\section{\sc The Piecewise Linear Approximation Model for Cumulative Hazard}\label{sec:Model}
In general, a PLA is a helpful tool for modelling data, avoiding strong parametric assumptions. In survival analysis, piecewise linear functions are used
extensively. 
Recently, Balakrishnan et al.~\cite{Balaetal2016} proposed a PLA-based model for the
hazard rate of a population with a cured proportion; see also the references therein. In
this article, we develop a PLA-based model for load-sharing systems with unknown
load-share rules. Specifically, we model the cumulative hazard functions of the component
lifetime distributions using PLAs. At each of the successive stages of component failures,
as the lifetime distributions of the remaining operating components change, a new PLA for
the cumulative hazard is used. The model can be suitably tuned by choosing the number of
linear pieces for the PLA at each stage of failure. The principal advantage of the
proposed PLA-based modelling approach is that it uses minimal model assumptions. 
   
Consider a $J$-component load-sharing system. Here, a $J$-component load-sharing system means a load-sharing system with $J$ components that are connected in parallel. Assume that the failed components of the system are not replaced or repaired. When the components fail one by one, after each failure the total load on the system gets redistributed over the remaining operational components. As a result the operational components experience a higher load than before. At the beginning when all components are operational, let $U_1^{(0)},\, U_2^{(0)},\, \ldots,\, U_J^{(0)}$ denote the latent lifetimes of the components, and $Y^{(0)}$ denote the system lifetime till the first component failure. Obviously,  
\begin{align*}
   Y^{(0)} = \min \left\{ U_1^{(0)},\, U_2^{(0)},\, \ldots,\, U_J^{(0)} \right\}.
\end{align*}
Similarly, for $j=1,\,2,\,\ldots,\,J-1$, let $Y^{(j)}$ denote the system lifetime between $j$-th and $(j+1)$-st component failures. Then, 
\begin{align*}
   Y^{(j)} = \min \left\{ U_1^{(j)},\, U_2^{(j)},\, \ldots,\, U_{J-j}^{(j)} \right\},
\end{align*}
where $U_{1}^{(j)},\, U_2^{(j)},\,\ldots\, U_{J-j}^{(j)}$ denote the latent lifetimes of
the operational components after the $j$-th component failure,
$j=1,\,2,\,\ldots,\,J-1$. For all values of $j$, $U_1^{(j)},\, \ldots,\,
U_{J-j}^{(j)}$ are assumed to be independent and identically distributed random variables.
It is further assumed that $\left\{U_\ell^{(j)}, \ell=1,\,2,\, \ldots,\, J-j;\, j=0,\,1,\, \ldots,\, J-1\right\}$ are independent random variables.

Let $h^{(j)}(\cdot)$ and $H^{(j)}(\cdot)$ denote the hazard rate (HR) and cumulative hazard
function (CHF), respectively, of the distribution of $U_{1}^{(j)},$ $j=0,\,1,\,2,\,\ldots,\,J-1$. Here, we assume that the HR $h^{(j)}\left( \cdot \right)$ is a non-decreasing function for all $j$. For $y>0$, the survival function (SF) of $Y^{(j)}$ is given by
\begin{align*}
   P\left( Y^{(j)}>y \right) 
   &= P\left[\text{min}\left\{ U_{1}^{(j)},\, U_{2}^{(j)},\, \ldots, U_{J-j}^{(j)}
   \right\} >y \right] = e^{-(J-j)H^{(j)}(y)}.
\end{align*}
Hence, for $y>0$, the cumulative distribution function (CDF) and probability density
function (PDF) of $Y^{(j)}$ are given by
\begin{align*}
   F^{(j)}(y) = 1-e^{-(J-j)H^{(j)}(y)} && \text{and} && f^{(j)}(y) = (J-j)h^{(j)}(y)\,
   e^{-(J-j)H^{(j)}(y)},
\end{align*}
respectively.

Now, suppose there are $n$ $J$-component load-sharing systems, and let $Y_{i}^{(j)}$ denote the system lifetime between $j$-th and $(j+1)$-st component failures for the $i$-th system, $i=1,\,2,\, \ldots,\, n$, $j=0,\,1,\, \ldots,\, J-1$. Suppose the observed values of $Y_1^{(j)},\, Y_{2}^{(j)},\, \ldots,\, Y_{n}^{(j)}$ are $y_{1}^{(j)},\, y_2^{(j)},\, \ldots,\, y_{n}^{(j)}$, respectively. Let, for $j=0,\,1,\, \ldots,\, J-1$,  $\xi^{(j)} = \left\{
\tau_0^{(j)},\, \tau_1^{(j)},\, \ldots,\, \tau_N^{(j)} \right\}$ denote a set of $N+1$
cut-points over the time scale $y_1^{(j)},\, \ldots,\, y_n^{(j)}$, with the restrictions that 
\begin{equation}
\tau_0^{(j)} < \tau_1^{(j)} < \tau_2^{(j)} < \ldots < \tau_{N}^{(j)},  \quad \tau_0
^{(j)} \le \min \left\{ y_1^{(j)},\, \ldots,\, y_n^{(j)} \right\} \textrm{ and } \tau_N
^{(j)} \ge \max \left\{ y_1^{(j)},\, \ldots,\, y_n^{(j)} \right\}. \nonumber
\end{equation}
Initially, $\xi^{(j)}$ is taken to be fixed and known. We discuss how to choose $\xi^{(j)}$ in a later section.

The proposed model approximates the CHF $H^{(j)}(\cdot)$ by a piecewise linear function
defined over intervals $[\tau_{k-1}^{(j)},\, \tau_{k}^{(j)})$, $k=1,\,2,\,\ldots,\,N$,
constructed by the consecutive cut points in $\xi^{(j)}$. Therefore, over the range $[\tau_0^{(0)},\,\tau_{N}^{(0)})$, the CHF $H^{(0)}(\cdot)$ is approximated by $\Lambda^{(0)}(\cdot)$, where
\begin{align}
   \Lambda^{(0)}(t) = \sum_{k=1}^{N} \left( a_k+b_kt \right)
   \boldsymbol{1}_{[\tau_{k-1}^{(0)}, \, \tau_k^{(0)})}(t), \label{eq:model0}
\end{align}
with $a_k$'s and $b_k$'s as real constants and
\begin{align*}
   \boldsymbol{1}_{A}(t) = \begin{cases}
      1 & \text{if } t\in A\\
      0 & \text{if } t\not\in A.
   \end{cases}
\end{align*}

One of the possible ways to extend the PLA beyond $\tau_{N}^{(0)}$ would be to extend the last line segment $a_N+b_Nt$ to $[\tau_N^{(0)},\,\infty)$. Therefore, the CHF corresponding to PLA over the range $[\tau_0^{(0)},\,\infty)$ is 
\begin{align*}
   \Lambda^{(0)}(t) = \sum_{k=1}^{N} \left( a_k+b_kt \right)
   \boldsymbol{1}_{[\tau_{k-1}^{(0)}, \, \tau_k^{(0)})}(t)+(a_N+b_N t) \boldsymbol{1}_{[\tau_{N}^{(0)}, \, \infty)}(t), 
\end{align*} with $\Lambda^{(0)}(\tau_0^{(0)})=0$.
We also assume that $\Lambda^{(0)}(\cdot)$ is a continuous function. As $\Lambda^{(0)}(\tau_0^{(0)})=0$, using the assumption of continuity, $a_i$'s can be expressed in terms of $b_i$'s as
follows:
\begin{align*}
   a_1=-b_1\tau_0^{(0)} && \text{and} && a_{k} = \sum_{\ell=1}^{k-1} \left( b_\ell - b_{\ell+1} \right)
   \tau_\ell^{(0)} +a_1 = \sum_{\ell=1}^{k-1} b_\ell \left( \tau_\ell^{(0)} -
   \tau_{\ell-1}^{(0)} \right) - b_k\tau_{k-1}^{(0)},
\end{align*} for $k = 1,\, 2,\,3,\,\ldots,\,N.$

Note that the above model can be equivalently described in terms of HRs. In this approach, $h^{(0)}(\cdot)$ over the range $[\tau_0^{(0)},\,\tau_{N}^{(0)})$ is approximated by a piecewise constant function $\lambda^{(0)}(\cdot)$, where  
\begin{align}
   \lambda^{(0)}(t) = \sum_{i=1}^{N} b_k \boldsymbol{1}_{[\tau^{(0)}_{k-1},\,
   \tau^{(0)}_{k})} \left( t \right). \label{eq:model1}
\end{align}

After failure of one or more components within the system, the direct impact of the increased load will be an increased HR for the operational components. To incorporate this information, after the failure of $j$ components of the system, we approximate $h^{(j)}(\cdot)$ over $[\tau_0^{(j)},\,\tau_{N}^{(j)})$, $j=1,\,2,\, \ldots,\, J-1$, using the piecewise constant function $\lambda^{(j)}(\cdot)$, where
\begin{align}
   \lambda^{(j)}(t) = \gamma_j \sum_{k=1}^{N} b_k \boldsymbol{1}_{[\tau^{(j)}_{k-1},\,
   \tau^{(j)}_{k})} \left( t \right), \label{eq:model2}
\end{align}
with 
$$1<\gamma_1<\gamma_2<\ldots< \gamma_{J-1}.$$
The PLAs to the CHFs, corresponding to the PLAs of the HRs given in Eq.\eqref{eq:model2} are given by
\begin{align}
   \Lambda^{(j)}(t) = \gamma_j \sum_{k=1}^{N}\left[ \sum_{\ell=1}^{k-1} b_\ell \left(
   \tau_\ell^{(j)}-\tau_{\ell-1}^{(j)} \right) + b_k \left( t-\tau_{k-1}^{(j)} \right)
   \right] \boldsymbol{1}_{[\tau_{k-1}^{(j)},\, \tau_k^{(j)})} \left( t \right). \label{eq:model3}
\end{align}
To meet the non-decreasing nature of the HR, we assume that $0<b_1<b_2<\ldots<b_N$. Note that the parameters $\gamma_1, \gamma_2, \ldots, \gamma_{J-1}$ reflect the load-share rule of increased HRs. We treat $\gamma_1, \gamma_2, \ldots, \gamma_{J-1}$ as unknown parameters, and estimate them from component failure data. It may be mentioned here that the PLA model can be interpreted as an approximation of the underlying lifetime distribution by several exponential models (with different rate parameters) over the ranges specified by the cut-points.

\section{\sc Likelihood Inference} \label{sec:Inf}
The parameters involved in the PLA-based model are estimated from the component failure data obtained from a set of load-sharing systems. The available data on component failures from $n$ $J$-component load-sharing systems is of the form 
$$
Data = \left\{ y_i^{(j)}\,:\, i=1,\,2,\,\ldots,\,n;\, j=0,\,1,\, \ldots,\, J-1 \right\},
$$
where $y_{i}^{(j)}$ is the observed system lifetime between $j$-th and $(j+1)$-st component failures for the $i$-th system. For $j=0,\,1,\,2,\,\ldots,\,J-1$, and $k=1,\,2,\,\ldots,\,N$, define 
\begin{align*}
   I_k^{(j)} = \left\{i:y_i^{(j)} \in \left[\tau_{k-1}^{(j)},\,\tau_k^{(j)}\right)
   \right\} \quad \textrm{and} \quad n_k^{(j)} = \vert I_k^{(j)} \vert. 
\end{align*}
Obviously, $\sum_{k=1}^{N}n_k^{(j)}=n$. The likelihood function for the PLA model is then given by  
\begin{align}
   L\left( \boldsymbol{\theta} \right) 
   &= \prod_{i=1}^{n}\prod_{j=0}^{J-1} \left[ (J-j)\gamma_j\sum_{k=1}^{N}b_k \boldsymbol{1}_{[\tau^{(j)}_{k-1},\,
   \tau^{(j)}_{k})} \left( y_i^{(j)} \right) 
   e^ {-(J-j)\gamma_j \left[\sum_{\ell=1}^{k-1} b_\ell \left(
   \tau_\ell^{(j)}-\tau_{\ell-1}^{(j)} \right) + b_k \left( y_i^{(j)}-\tau_{k-1}^{(j)} \right)
   \right] } \right] , \label{eq:lik}
\end{align}
where $\gamma_0=1$ and $\boldsymbol{\theta}
= \left(\gamma_1,\, \gamma_2,\, \ldots,\, \gamma_{J-1}, b_1,\, b_2,\, \ldots,\, b_N \right)^\prime$ is the vector of parameters. The corresponding log-likelihood function, ignoring additive constant, can be expressed as
\begin{align}
   l\left( \boldsymbol{\theta} \right) 
   &= \sum_{k=1}^{N} \left[ \left( \sum_{j=0}^{J-1} n_k^{(j)} \right) \ln b_k -
   \left( \sum_{j=0}^{J-1} (J-j)\gamma_j T_k^{(j)} \right) b_k \right] + n \sum_{j=0}^{J-1}
   \ln\gamma_j, \label{eq:loglik}
\end{align}
where
\begin{align*}
   T_k^{(j)} &= \sum_{i\in I_k^{(j)}}\left( y_i^{(j)}-\tau_{k-1}^{(j)}\right) + \left( n-\sum_{\ell=1}^{k} n_\ell^{(j)} \right) \left(
   \tau_k^{(j)}-\tau_{k-1}^{(j)} \right),
\end{align*}
for $k=1,\,2,\,\ldots,\,N;\, j=0,\,1,\, \ldots,\, J-1$. Equating partial derivative of the
log-likelihood function in Eq.\eqref{eq:loglik} with respect to $b_k$ to zero, we can
express $b_k$ in terms of the load-share parameters $\boldsymbol{\gamma}= \left(
\gamma_1,\, \gamma_2,\, \ldots,\, \gamma_{J-1} \right)$ as
\begin{align}
   b_k = b_{k}\left( \boldsymbol{\gamma} \right) =
   \frac{\displaystyle \sum_{j=0}^{J-1} n_k^{(j)}}{\displaystyle \sum_{j=0}^{J-1}
   (J-j)\gamma_j T_k^{(j)}}, \quad k = 1,...,N. \label{eq:slope-mles}
\end{align}
Substituting $b_k(\boldsymbol{\gamma})$ from Eq.\eqref{eq:slope-mles} in Eq.\eqref{eq:loglik}, the profile log-likelihood in $\boldsymbol{\gamma}$, ignoring
additive constant, is obtained as
\begin{align}
\tilde l\left( \boldsymbol{\gamma} \right) 
&= \sum_{k=1}^{N} \left[ \left( \sum_{j=0}^{J-1} n_k^{(j)} \right) \left\{\ln \left( \sum_{j=0}^{J-1} n_k^{(j)} \right)-\ln \left( \sum_{j=0}^{J-1} (J-j)\gamma_j T_k^{(j)} \right)\right\} \right] + n \sum_{j=0}^{J-1}
\ln\gamma_j. \label{eq:prof_loglik}
\end{align}
For optimizing the profile log-likelihood $\tilde l\left( \boldsymbol{\gamma} \right)$ in $\boldsymbol{\gamma}$, any routine maximizer of a standard statistical software may be used. Once the MLEs $\widehat{\gamma}_1,\, \widehat{\gamma}_2,\,\ldots,\, \widehat{\gamma}_{J-1}$ of $\gamma_1,\,\gamma_2,\, \ldots,\, \gamma_{J-1}$ are
obtained by numerical optimization of $\tilde l\left( \boldsymbol{\gamma} \right)$,
they can be plugged into Eq.\eqref{eq:slope-mles} to get MLEs of $b_k$ as
$$
\widehat{b}_k = b_k\left( \widehat{\gamma}_1,\, \ldots,\,
\widehat{\gamma}_{J-1} \right), \quad k=1,\,2,\, \ldots,\, N.
$$ 

\subsection{\sc A special case: two-component load-sharing systems} \label{subsec:2-comp-mle}
For analysing data from two-component load-sharing systems, if two linear pieces are used in the PLA-based model, MLEs can be derived analytically and explicitly. Consider the case when $J=2$ and $N=2$. In this case, the log-likelihood function simplifies to 
\begin{align}
   l\left( \boldsymbol{\theta} \right) 
   &= \sum_{k=1}^{2} \left[ \left( \sum_{j=0}^{1} n_k^{(j)} \right) \ln b_k -
   \left( \sum_{j=0}^{1} (2-j)\gamma_j T_k^{(j)} \right) b_k \right] + n \sum_{j=0}^{1}
   \ln\gamma_j, \label{eq:loglik2}
\end{align}
with 
\begin{align*}
   T_k^{(j)} &= \sum_{i\in I_k^{(j)}}\left( y_i^{(j)}-\tau_{k-1}^{(j)}\right) + \left( n-\sum_{\ell=1}^{k} n_\ell^{(j)} \right) \left(
   \tau_k^{(j)}-\tau_{k-1}^{(j)} \right),
\end{align*}
for $k=1, 2$, $j=0, 1$ and $\gamma_0=1$. Here, $\boldsymbol{\theta} = (\gamma_1,b_1, b_2)$. 

Equating $\frac{\partial l\left(\boldsymbol{\theta}\right)}{\partial b_1}$ and $\frac{\partial l(\boldsymbol{\theta})}{\partial b_2}$ to zero, we get 
\begin{align}
   b_1= \frac{n_1^{(0)}+n_1^{(1)}}{2T_1^{(0)}+\gamma_1 T_1^{(1)}}\label{eq:b1-sc}
\end{align}
\begin{align}
   b_2=\frac{n_2^{(0)}+n_2^{(1)}}{2T_2^{(0)}+\gamma_1 T_2^{(1)}}.\label{eq:b2-sc}
\end{align}
Equating $\frac{\partial l\left(\boldsymbol{\theta}\right)}{\partial \gamma_1}$ to zero gives
\begin{align}
   \gamma_1=T_1^{(1)}b_1+T_2^{(1)}b_2, \label{eq:gamma-sc}
\end{align}
in which, substituting $b_1$ and $b_2$ from Eqs.\eqref{eq:b1-sc} and \eqref{eq:b2-sc}, a quadratic equation in $\gamma_1$ is obtained as follows 
\begin{align}
   Q(\gamma_1) = n\gamma_1^2B_{0,12}+2\gamma_1 \left\{\left(n_1^{(0)}+n_1^{(1)}-n\right)B_{2,1}+\left(n_2^{(0)}+n_2^{(1)}-n\right)B_{1,2}\right\}-4nB_{12,0}=0, \label{eq:gamma1-Q}
\end{align}
with $B_{0,12}=T_1^{(1)}T_2^{(1)}$, $B_{1,2}=T_1^{(0)}T_2^{(1)}$, $B_{2,1}=T_2^{(0)}T_1^{(1)}$ and $B_{12,0}=T_1^{(0)}T_2^{(0)}.$
Solving $Q(\gamma_1) = 0$, we have two values of $\gamma_1$ from which we choose the suitable one, and then from equations \eqref{eq:b1-sc} and \eqref{eq:b2-sc} we get the MLEs of $b_1$ and $b_2$, respectively.

\subsection{\sc Confidence Intervals} \label{subsec:CI}
As discussed above, the MLEs for the parameters of the PLA-based model are not available in explicit form in general, except for the special case of two-component load-sharing systems considered in Section \ref{subsec:2-comp-mle}. As a result, exact confidence
intervals for the model parameters cannot be obtained. Asymptotic confidence intervals may be constructed in two possible ways: by using the Fisher information matrix, and by applying a bootstrap-based technique.   

\subsubsection{\sc CIs using Fisher information matrix}
Using the asymptotic properties of the MLEs, it can be shown that for large sample size $n$, the distribution of $\sqrt{n}(\widehat{\boldsymbol \theta} - \boldsymbol \theta)$ is approximated by a multi-variate normal distribution $\boldsymbol N(\boldsymbol 0, \mathbb{
I}^{-1}(\widehat{\boldsymbol \theta}))$, where the dimension of the multi-variate normal
distribution is same as that of the parameter vector $\boldsymbol \theta$, and the asymptotic variance-covariance matrix $\mathbb{I}^{-1}(\boldsymbol \theta)$ is the inverse of the Fisher information matrix $\mathbb{I}(\boldsymbol \theta)$, evaluated at the MLE $\widehat{\boldsymbol \theta}$. The Fisher information matrix $\mathbb{I}(\boldsymbol \theta)$ is defined as the expected value of the observed information matrix $\mathbb{J}(\boldsymbol \theta)$ which is calculated from the negative of the second-order derivatives of the log-likelihood function. That is,  $\mathbb{I}(\boldsymbol{\theta})=\text{E}(\mathbb{J}(\boldsymbol{\theta}))$, where $\mathbb{ J}(\boldsymbol \theta) = -\nabla^2(\log L(\boldsymbol \theta))$. In situations where analytical calculation of the Fisher information is difficult or intractable, it may be either replaced by the observed information matrix, or may be calculated by simulations. 

From the asymptotic variance-covariance matrix $\mathbb{I}^{-1}(\boldsymbol \theta)$, individual asymptotic variances of the MLEs can be pulled out, and asymptotic confidence intervals can be constructed. For example, corresponding to the MLE $\widehat{\gamma}_1$ using the asymptotic variance $\widehat{Var(\hat{\gamma}_1)}$ obtained from $\mathbb{I}^{-1}(\boldsymbol \theta)$, asymptotic confidence intervals for $\gamma_1$ can be constructed as: $$\left(\widehat{\gamma}_1 - z_{\alpha/2} \sqrt{\widehat{Var(\hat{\gamma}_1)}}, \widehat{\gamma}_1 + z_{\alpha/2} \sqrt{\widehat{Var(\hat{\gamma}_1)}}\right),$$
where $z_{\alpha}$ is the $100(1-\alpha)$\% point of the standard normal distribution. 

\subsubsection*{\sc Special case: two-component load-sharing systems}
For the special case of two-component load-sharing systems considered in Section \ref{subsec:2-comp-mle}, the Fisher information matrix can be worked out explicitly. In this case,
\begin{align*}
   \mathbb{J}(\boldsymbol{\theta})=-\begin{pmatrix}
      \frac{\partial^2 l(\boldsymbol{\theta})}{\partial \gamma_1^2} & \frac{\partial^2 l(\boldsymbol{\theta})}{\partial \gamma_1 \partial b_1} & \frac{\partial^2 l(\boldsymbol{\theta})}{\partial \gamma_1 \partial b_2}\\\\ 
      \frac{\partial^2 l(\boldsymbol{\theta})}{\partial b_1 \partial \gamma_1} & \frac{\partial^2 l(\boldsymbol{\theta})}{\partial b_1^2} & \frac{\partial^2 l(\boldsymbol{\theta})}{\partial b_1 \partial b_2}\\\\
      \frac{\partial^2 l(\boldsymbol{\theta})}{\partial b_2 \partial \gamma_1} & \frac{\partial^2 l(\boldsymbol{\theta})}{\partial b_2 \partial b_1} & \frac{\partial^2 l(\boldsymbol{\theta})}{\partial b_2^2} 
   \end{pmatrix} = -\begin{pmatrix}
      -\frac{n}{\gamma_1^2} & -T_1^{(1)} & -T_2^{(1)}\\
      -T_1^{(1)} & -\frac{n_1^{(0)}+n_1^{(1)}}{{b_1}^2} & 0\\
      -T_2^{(1)} & 0 & -\frac{n_2^{(0)}+n_2^{(1)}}{{b_2}^2}
  \end{pmatrix}.
\end{align*}

  Hence, the Fisher information matrix is 
  \scriptsize
   \begin{align*}
      \mathbb{I}(\boldsymbol{\theta})=\begin{pmatrix}
   \frac{n}{\gamma_1^2} & E\left(\displaystyle\sum_{i \in I_1^{(1)}}Y_i^{(1)}\right)+E\left(N_2^{(1)}\right)\tau_1^{(1)} & E\left(\displaystyle\sum_{i \in I_1^{(1)}}Y_i^{(1)}\right)-E\left(N_2^{(1)}\right)\tau_1^{(1)}\\\\
  E\left(\displaystyle\sum_{i \in I_1^{(1)}}Y_i^{(1)}\right)+E\left(N_2^{(1)}\right)\tau_1^{(1)} & \frac{E\left(N_1^{(0)}\right)+E\left(N_1^{(1)}\right)}{b_1^2} & 0\\\\
 E\left(\displaystyle\sum_{i \in I_1^{(1)}}Y_i^{(1)}\right)-E\left(N_2^{(1)}\right)\tau_1^{(1)} & 0 & \frac{E\left(N_2^{(0)}\right)+E\left(N_2^{(1)}\right)}{b_2^2} 
 \end{pmatrix},
   \end{align*}
   \normalsize
   where $N_k^{(j)}$ is the number of $Y_i^{(j)}$ in $[\tau_{k-1}^{(j)}, \tau_k^{(j)})$, $k=1,2$, $j=0,1$, $i=1,...,n$. An outline of calculations of the relevant expectations for the Fisher information matrix is given in Appendix A. The inverse of the Fisher information matrix is obtained as
   \begin{align*}
      \left\{\mathbb{I}^{-1}(\boldsymbol{\theta})\right\}=\frac{1}{|\mathbb{I}(\boldsymbol{\theta})|}\begin{pmatrix}
         A_{11}(\boldsymbol{\theta}) & -A_{12}(\boldsymbol{\theta}) & A_{13}(\boldsymbol{\theta})\\
         -A_{21}(\boldsymbol{\theta}) & A_{22}(\boldsymbol{\theta}) & -A_{23}(\boldsymbol{\theta})\\
         A_{31}(\boldsymbol{\theta}) & -A_{32}(\boldsymbol{\theta}) & A_{33}(\boldsymbol{\theta})
      \end{pmatrix},
   \end{align*}  
   where the determinant of $\mathbb{I}(\boldsymbol{\theta})$ is
   \begin{eqnarray}
      &|\mathbb{I}(\boldsymbol{\theta})| &= \frac{n\left\{2-\left(e^{-2b_1\tau_1^{(0)}}+e^{-\gamma_1 b_1\tau_1^{(1)}}\right)\right\}\left(e^{-2b_1\tau_1^{(0)}}+e^{-\gamma_1 b_1\tau_1^{(1)}}\right)}{\gamma_1^2b_1^2b_2^2} \nonumber \\
      &&-\frac{e^{-2\gamma_1 b_1\tau_1^{(1)}}\left(\frac{1}{\gamma_1 b_2}\right)^2\left\{2-\left(e^{-2b_1\tau_1^{(0)}}+e^{-\gamma_1 b_1\tau_1^{(1)}}\right)\right\}}{b_1^2} \nonumber \\
      &&-\frac{\left\{\frac{1}{\gamma_1 b_1}\left[1-(1+\gamma_1 b_1 \tau_1^{(1)})e^{-\gamma_1 b_1\tau_1^{(1)}}\right]+\tau_1^{(1)}e^{-\gamma_1 b_1\tau_1^{(1)}}\right\}^2\left(e^{-2b_1\tau_1^{(0)}}+e^{-\gamma_1 b_1\tau_1^{(1)}}\right)}{b_2^2}, \nonumber
   \end{eqnarray}
   \begin{align*}
      A_{11}(\boldsymbol{\theta})=\frac{\left\{2-\left(e^{-2b_1\tau_1^{(0)}}+e^{-\gamma_1 b_1\tau_1^{(1)}}\right)\right\}\left(e^{-2b_1\tau_1^{(0)}}+e^{-\gamma_1 b_1\tau_1^{(1)}}\right)}{b_1^2b_2^2},
   \end{align*}
   \begin{align*}
      A_{22}(\boldsymbol{\theta})=
      &\frac{n\left(e^{-2b_1\tau_1^{(0)}}+e^{-\gamma_1
      b_1\tau_1^{(1)}}\right)}{\gamma_1^2b_2^2}-e^{-2\gamma_1
      b_1\tau_1^{(1)}}\left(\frac{1}{\gamma_1 b_2}\right)^2,\\
      A_{33}(\boldsymbol{\theta})=
      &\frac{n\left\{2-\left(e^{-2b_1\tau_1^{(0)}}+e^{-\gamma_1
      b_1\tau_1^{(1)}}\right)\right\}}{\gamma_1^2b_1^2}\\
      & -\left\{\frac{1}{\gamma_1 b_1}\left[1-(1+\gamma_1 b_1 \tau_1^{(1)})e^{-\gamma_1 b_1\tau_1^{(1)}}\right]+\tau_1^{(1)}e^{-\gamma_1 b_1\tau_1^{(1)}}\right\}^2,
   \end{align*}
   \begin{align*}
      A_{12}(\boldsymbol{\theta})=A_{21}(\boldsymbol{\theta})=
      & \frac{\left\{\frac{1}{\gamma_1
      b_1}\left[1-(1+\gamma_1 b_1 \tau_1^{(1)})e^{-\gamma_1
      b_1\tau_1^{(1)}}\right]+\tau_1^{(1)}e^{-\gamma_1
      b_1\tau_1^{(1)}}\right\}\left(e^{-2b_1\tau_1^{(0)}}+e^{-\gamma_1
      b_1\tau_1^{(1)}}\right)}{b_2^2},\\
      A_{13}(\boldsymbol{\theta})=A_{31}(\boldsymbol{\theta})=
      &-\frac{e^{-\gamma_1 b_1\tau_1^{(1)}}\left(\frac{1}{\gamma_1
      b_2}\right)\left\{2-\left(e^{-2b_1\tau_1^{(0)}}+e^{-\gamma_1
      b_1\tau_1^{(1)}}\right)\right\}}{b_1^2},\\
      A_{23}(\boldsymbol{\theta})=A_{32}(\boldsymbol{\theta})=
      &-\frac{\left\{\left[1-(1+\gamma_1 b_1 \tau_1^{(1)})e^{-\gamma_1
      b_1\tau_1^{(1)}}\right]+\gamma_1 b_1\tau_1^{(1)}e^{-\gamma_1
      b_1\tau_1^{(1)}}\right\}e^{-\gamma_1 b_1\tau_1^{(1)}}}{\gamma_1^2b_1b_2}.
   \end{align*}

Evaluating $\mathbb{I}^{-1}(\boldsymbol{\theta})$ at the MLE $\widehat{\boldsymbol \theta}$, the asymptotic variance-covariance matrix of the MLEs is obtained. Hence, $100(1-\alpha)$\% asymptotic confidence intervals for $\gamma_1$, $b_1$, and $b_2$ are obtained as 
\bigg($\widehat{\gamma}_1-z_{\alpha/2} \sqrt{\frac{A_{11}(\boldsymbol{\hat{\theta}})}{|\mathbb{I}(\boldsymbol{\hat{\theta}})|}}, \widehat{\gamma}_1+z_{\alpha/2} \sqrt{\frac{A_{11}(\boldsymbol{\hat{\theta}})}{|\mathbb{I}(\boldsymbol{\hat{\theta}})|}}$\bigg),  \bigg($\widehat{b_1}-z_{\alpha/2} \sqrt{\frac{A_{22}(\boldsymbol{\hat{\theta}})}{|\mathbb{I}(\boldsymbol{\hat{\theta}})|}}, \widehat{b_1}+z_{\alpha/2} \sqrt{\frac{A_{22}(\boldsymbol{\hat{\theta}})}{|\mathbb{I}(\boldsymbol{\hat{\theta}})|}}$\bigg), and \bigg($\widehat{b_2}-z_{\alpha/2} \sqrt{\frac{A_{33}(\boldsymbol{\hat{\theta}})}{|\mathbb{I}(\boldsymbol{\hat{\theta}})|}}, \widehat{b_2}+z_{\alpha/2} \sqrt{\frac{A_{33}(\boldsymbol{\hat{\theta}})}{|\mathbb{I}(\boldsymbol{\hat{\theta}})|}}$\bigg), respectively.
 
\subsubsection{\sc Bootstrap confidence intervals}
Using the MLE $\widehat{\boldsymbol \theta}$, $B$ bootstrap samples can be obtained in the
same sampling framework; let $\widehat{\boldsymbol
\theta}_s^*=\left(\widehat{\gamma}_{1s}^*, \widehat{b}_{1s}^*, \widehat{b}_{2s}^*\right)$
denote the bootstrap estimates, $s=1,...,B$. Bootstrap bias and standard error are defined as 
$$
bias_{b}(\widehat{\gamma}_1) = \overline{\widehat{\gamma_1^*}} - \widehat{\gamma}_1, \quad  bias_{b}(\widehat{b}_1) = \overline{\widehat{b_1^*}} - \widehat{b}_1, \quad bias_{b}(\widehat{b}_2) = \overline{\widehat{b_2^*}} - \widehat{b}_2
$$ 
and 
\scriptsize
$$
SE_{b}(\widehat{\gamma}_1)=\sqrt{\frac{1}{B-1}\sum_{s=1}^{B}\left(\widehat{\gamma}_{1s}^*-\overline{\widehat{\gamma_1^*}}\right)^2}, SE_{b}(\widehat{b}_1)=\sqrt{\frac{1}{B-1}\sum_{s=1}^{B}\left(\widehat{b}_{1s}^*-\overline{\widehat{b_1^*}}\right)^2}, SE_{b}(\widehat{b}_2)=\sqrt{\frac{1}{B-1}\sum_{s=1}^{B}\left(\widehat{b}_{2s}^*-\overline{\widehat{b_2^*}}\right)^2},
$$
\normalsize
where 
$$
\overline{\widehat{\gamma_1^*}} = \frac{1}{B}\sum_{s=1}^{B}\widehat{\gamma}_{1s}^*, \quad \overline{\widehat{b_1^*}} = \frac{1}{B}\sum_{s=1}^{B}\widehat{b}_{1s}^*, \quad \overline{\widehat{b_2^*}} = \frac{1}{B}\sum_{s=1}^{B}\widehat{b}_{2s}^*. 
$$
Finally, a $100(1-\alpha)$\% bootstrap confidence interval for $\gamma_1$ can be calculated as
\begin{align*}
\left(\widehat{\gamma}_1 - bias_b(\widehat{\gamma}_1) - z_{\alpha/2}SE_{b}(\widehat{\gamma}_1), \widehat{\gamma}_1 - bias_b(\widehat{\gamma}_1) + z_{\alpha/2}SE_{b}(\widehat{\gamma}_1)\right).
\end{align*}
Bootstrap confidence intervals for $b_1$ and $b_2$ can be calculated similarly. 

For percentile bootstrap confidence intervals for, say $\gamma_1$, the bootstrap estimates of $\widehat{\gamma}_1$ are first ordered in terms of magnitude: 
$$
\widehat{\gamma}_{1(1)}^* < \widehat{\gamma}_{1(2)}^* < ... < \widehat{\gamma}_{1(B)}^*.
$$ 
Then, a $100(1-\alpha)$\% percentile bootstrap confidence interval for $\gamma_1$ is $\left( \widehat{\gamma}_{1([\frac{\alpha B}{2}])}^*,\, 
\widehat{\gamma}_{1([(1-\frac{\alpha}{2})B])}^* \right)$. Similarly, percentile bootstrap
confidence intervals can be calculated for $b_1$ and $b_2$. 

\subsection{\sc Choice of Cut Points}
The number and position of the cut-points for constructing the PLA-based model need to be
suitably chosen, so that the model can closely approximate the underlying CHF, but avoid
overfitting. A large number of cut points would provide a close local approximation to the
underlying CHF. However, apart from being computationally expensive, a close local
approximation may also lead to overfitting in which case it would be difficult to use the
PLA-based model to predict future failures of components or systems. 

One of the possible ways to choose the number and position of the cut-points is by looking
at the plot of the nonparametric estimator of CHF. From such a plot, observing the areas where the nonparametric estimate changes significantly, one can determine the positions and number of cut-points.  

More objectively, one can choose the positions of a given number of cut-points by maximizing the log-likelihood function. For example, for three cut-points ($N=2$), the natural choice for $\tau_0^{(j)}$ is $\min\left\{ y_1^{(j)},\,
\ldots,\, y_n^{(j)} \right\}$ and $\tau_2^{(j)}$ is $\max\left\{ y_1^{(j)},\,
\ldots,\, y_n^{(j)} \right\}$. Now to choose the position of  $\tau_1^{(j)}$, one may take $\tau_1^{(j)}$ equal to different sample quantiles of $\left\{ y_1^{(j)},\,
\ldots,\, y_n^{(j)} \right\}$ and choose one that provides the maximum value of log-likelihood function evaluated at MLE. This process can be expressed as an algorithm as follows.\\
\textbf{Algorithm:}
\begin{itemize}
    \item \textbf{Step 1:} Fix $0<p_1<p_2<1$.
    \item \textbf{Step 2:} Find the number of $y_1^{(j)},\,
\ldots,\, y_n^{(j)}$ that are between $p_1$-th and $p_2$-th sample quantiles of $\left\{ y_1^{(j)},\,
\ldots,\, y_n^{(j)} \right\}$. Denote this number by $l$. Note that $l$ does not depend on $j=0,1,\ldots, J-1$.
    \item \textbf{Step 3:} Set $a_{j1}=p_1$-th quantile of $\left\{ y_1^{(j)},\,
\ldots,\, y_n^{(j)} \right\}, ~j=0,1,\ldots, J-1$.
    \item \textbf{Step 4:} Set $LL_1$= the value of log-likelihood function evaluated at MLE taking $\tau_1^{(j)}=a_{j1}, ~j=0,1,\ldots, J-1$.
    \item \textbf{Step 5:} Set $a_{j2}=\min \left\{y_i^{(j)}>a_{j1}; i=1, 2, \ldots, n\right\}, ~j=0,1,\ldots, J-1$.
    \item \textbf{Step 6:} Set $LL_2$= the value of log-likelihood function evaluated at MLE taking $\tau_1^{(j)}=a_{j2}, ~j=0,1,\ldots, J-1$.
    \item \textbf{Step 7:} Repeat the steps 5 and 6 to obtain $LL_1, LL_2, \ldots, LL_l$.
    \item \textbf{Step 8:} Set $k^*=\underset{1\le k \le l}{\arg \max}~ LL_k$.
    \item \textbf{Step 9:} The final cut points are $\tau_1^{(j)}=a_{jk^*}, ~j=0,1,\ldots, J-1$.
\end{itemize}

\section{\sc Estimation of various reliability characteristics}\label{sec:RC}
The final goal of fitting a model to load-sharing data, naturally, is accurate estimation
of reliability characteristics of load-sharing systems. As the PLA-based model provides a
good fit to load-sharing data due to the model's flexible nature, it is natural that the
important reliability characteristics of load-sharing systems can also be estimated quite
accurately under this model. In this section, we develop estimates of reliability
characteristics such as the quantile function, MTTF, RMT, and MRT of load-sharing systems
under the PLA-based model. Details of these derivations are given in Appendix~B for
interested readers. 

Under the PLA-based model, the quantile function of $Y^{(j)}$ which is the system lifetime between the $j$-th and $(j+1)$-st component failures, $j=0,...,J-1$, is given by 
\begin{align*}
 \eta(p)=\inf \left\{y\in \mathbb{R}: G^{(j)}(y)\ge p \right\}, \quad 0 < p < 1,
\end{align*} where $G^{(j)}(y)=1-e^{-(J-j)\Lambda^{(j)}(y)}.$
Using the expression of $\Lambda^{(j)}(y)$ given in Section \ref{sec:Model}, it is possible to work out an explicit formula for the quantile function $\eta(p)$, as follows: 
\begin{align*}
\eta(p)= 
\begin{cases}
\tau_{k-1}^{(j)}-\frac{\log(1-p)}{(J-j)\gamma_jb_k}-\frac{1}{b_k}\cdot \displaystyle\sum_{\ell=1}^{k-1}b_\ell(\tau_\ell^{(j)}-\tau_{\ell-1}^{(j)}), \text{ if } p\in \left[G^{(j)}(\tau_{k-1}^{(j)}), G^{(j)}(\tau_{k}^{(j)})\right),\\ \hspace{7.5cm}\text{ for } k=1, 2, \ldots, N.\\
\tau_{N-1}^{(j)}-\frac{\log(1-p)}{(J-j)\gamma_jb_N}-\frac{1}{b_N}\cdot \displaystyle\sum_{\ell=1}^{N-1}b_\ell(\tau_\ell^{(j)}-\tau_{\ell-1}^{(j)}), \text{ if } p\in \left[G^{(j)}(\tau_{N}^{(j)}), 1\right).
\end{cases}
\end{align*} 

The mean time to failure or MTTF of a load-sharing system is the expected time the system operates till its failure. Let $T$ denote the system failure time; then, $T=\sum_{j=0}^{J-1}Y^{(j)}$. 
The MTTF of a load-sharing system under the PLA-based model 
is given by
\begin{align*}
   E(T)=\sum_{j=0}^{J-1} \sum_{s=1}^{N}  \left\{\frac{e^{-\kappa_{j,s-1}}-e^{-\kappa_{j,s}}}{(J-j)\gamma_j b_\ell}\right\},  
\end{align*}
where
\begin{align*}
   \kappa_{j,s} = (J-j)\gamma_j \sum_{\ell=1}^{s} b_\ell
   \left(\tau_\ell^{(j)}-\tau_{\ell-1}^{(j)}\right).
\end{align*}

Reliability at a mission time or RMT of a system is the probability that the system will
operate till a desired time $t_0$; it is calculated as the survival probability of the
system at time $t_0$, i.e.,
$S(t_0)=P(T>t_0)=P\left(\displaystyle\sum_{j=0}^{J-1}Y^{(j)}>t_0\right)$. An explicit
expression for RMT may be derived by using the distribution of the system lifetime $T$. 

However, as $Y^{(j)}$s, $j=0,...,J-1$ are independent but not identically distributed, it is difficult to obtain an explicit expression for the distribution of the system lifetime $T$, where $T=\sum_{j=0}^{J-1}Y^{(j)}$. It is evident from the moment generating function $\phi_T(t)$ of $T$, which, under the PLA-based model, is given by  
\begin{align*}
   \phi_T(t) = \prod_{j=0}^{J-1}\sum_{s=1}^{N}\frac{(J-j)b_s\gamma_j}{(J-j) b_s\gamma_j-t}
   \left( e^{t\tau_{s-1}^{(j)}-\kappa_{j,s-1}} - e^{t\tau_s^{(j)}-\kappa_{j,s}}\right)
   \qquad \text{if } t<\gamma_1b_N,
\end{align*}
where
\begin{align*}
   \kappa_{j,s} = (J-j)\gamma_j \sum_{\ell=1}^{s} b_\ell
   \left(\tau_\ell^{(j)}-\tau_{\ell-1}^{(j)}\right).
\end{align*}
From here, it is clear that it is difficult to find the RMT analytically under this model.
However, for this model, RMT can be estimated using Monte Carlo simulations.
For a Monte Carlo estimate of the RMT at a pre-specified time $t_0$, one needs to generate
$R$ data points $t_i$, $i=1,\,2,\,\ldots,\,R$, as realisations of the system lifetime $T$,
and find $\frac{R(t_0)}{R}$, where $R(t_0)$ is the number of realisations of the system
lifetime that exceed $t_0$. For a reasonably good estimate of RMT, a large value of $R$
should be used.

The mean residual time or MRT of a system is the expected additional time the system will
survive if it has already survived a given time $t$. That is, 
$$
\textrm{MRT}(t) = E(T-t|T>t) = \int_{t}^{\infty}sf_{T|T>t}(s)ds-t.
$$
Therefore, analytical derivation of MRT requires the truncated distribution of the system
lifetime $T$, and it is difficult to obtain the truncated distribution of $T$ in this
case. Instead, an estimate of the MRT can be given using Monte Carlo simulations. We
generate $R$ data points $t_i^*$, $i=1,\,2,\,\ldots,\,R$, as realisations of the truncated
lifetime $T|T>t$, and a Monte Carlo estimate of the MRT for load-sharing systems under the
PLA-based model is then given by 
$$
\widehat{\textrm{MRT}}(t) = \frac{\displaystyle\sum_{i=1}^{R}t_i^*}{R} - t.
$$

\section{\sc Data Analysis} \label{sec:dataana}
In this section, we present an illustrative example using data from load-sharing systems comprising of two components. Very recently, this data have been analysed by Sutar and Naik-Nimbalkar~\cite{SN2014}, Asha et~al.~\cite{ARR2018} and Franco et~al.~\cite{FVK2020}. The data consist of information on component lifetimes of 18 two-component load-sharing systems. Each system is a parallel combination of two motors - ``A'' and ``B''. When both motors A and B are in working condition, the total load on the system is shared between them. When one of the motors fails, the entire load goes to the operational motor. 

\begin{figure}[!ht]
	\begin{subfigure}{.5\textwidth}
		\centering
		\includegraphics[width=\linewidth]{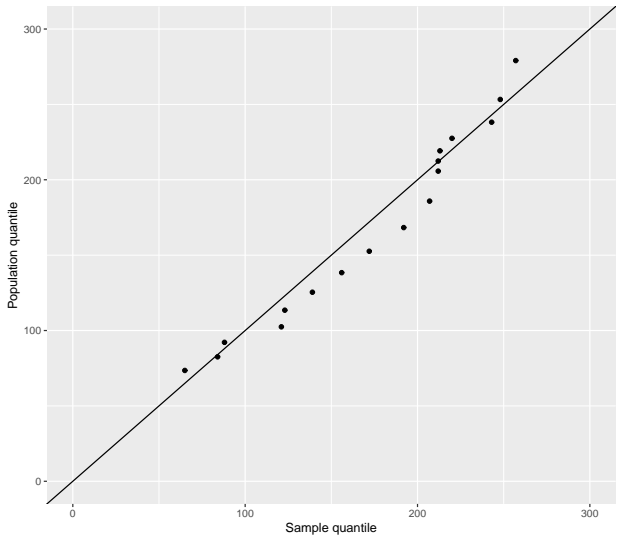}  
		\caption{Q-Q plot for $Y^{(0)}$}
		\label{fig:sub-qq0}
	\end{subfigure}
	\begin{subfigure}{.5\textwidth}
		\centering
		\includegraphics[width=\linewidth]{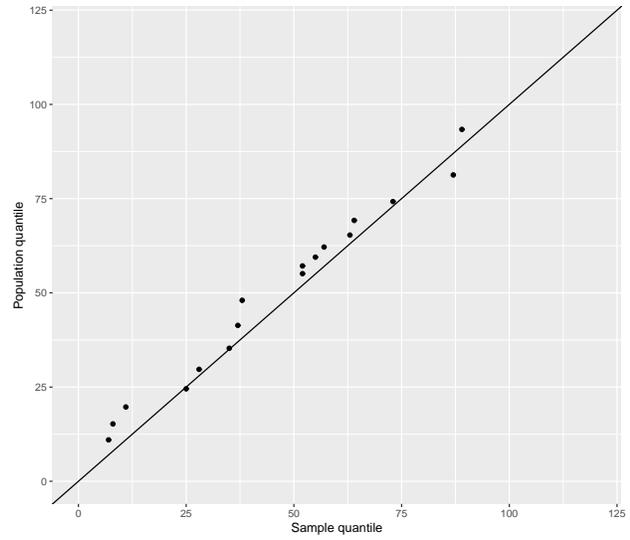}  
		\caption{Q-Q plot for $Y^{(1)}$}
		\label{fig:sub-qq1}
	\end{subfigure}
	\caption{Q-Q plots}
	\label{fig:qq}
\end{figure}

\begin{figure}[!ht]
	\begin{subfigure}{.5\textwidth}
		\centering
		\includegraphics[width=\linewidth]{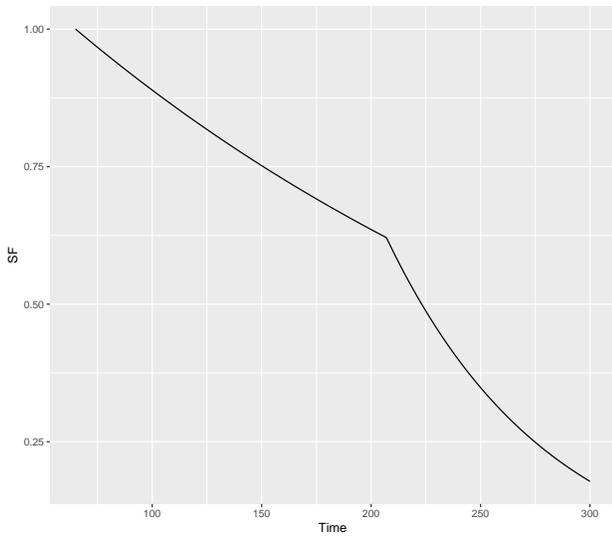}  
		\caption{Plot of SF for $Y^{(0)}$}
		\label{fig:sub-first2}
	\end{subfigure}
	\begin{subfigure}{.5\textwidth}
		\centering
		\includegraphics[width=\linewidth]{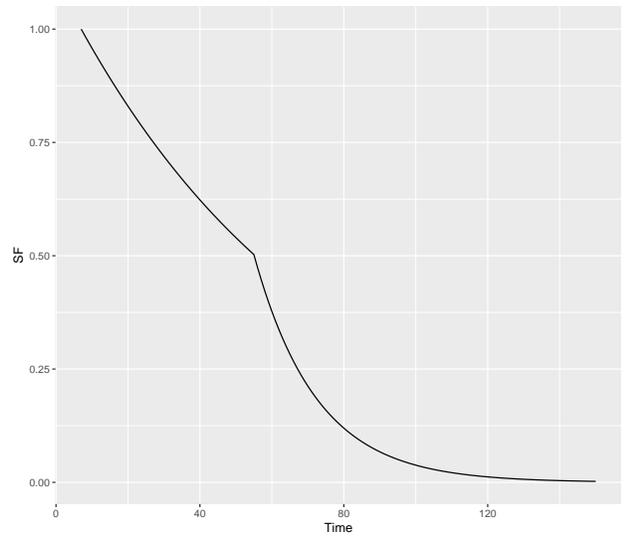}  
		\caption{Plot of SF for $Y^{(1)}$}
		\label{fig:sub-second2}
	\end{subfigure}
	\caption{Plots of SFs}
	\label{fig:fig2}
\end{figure}

\begin{figure}[!ht]
	\begin{subfigure}{.5\textwidth}
		\centering
		\includegraphics[width=\linewidth]{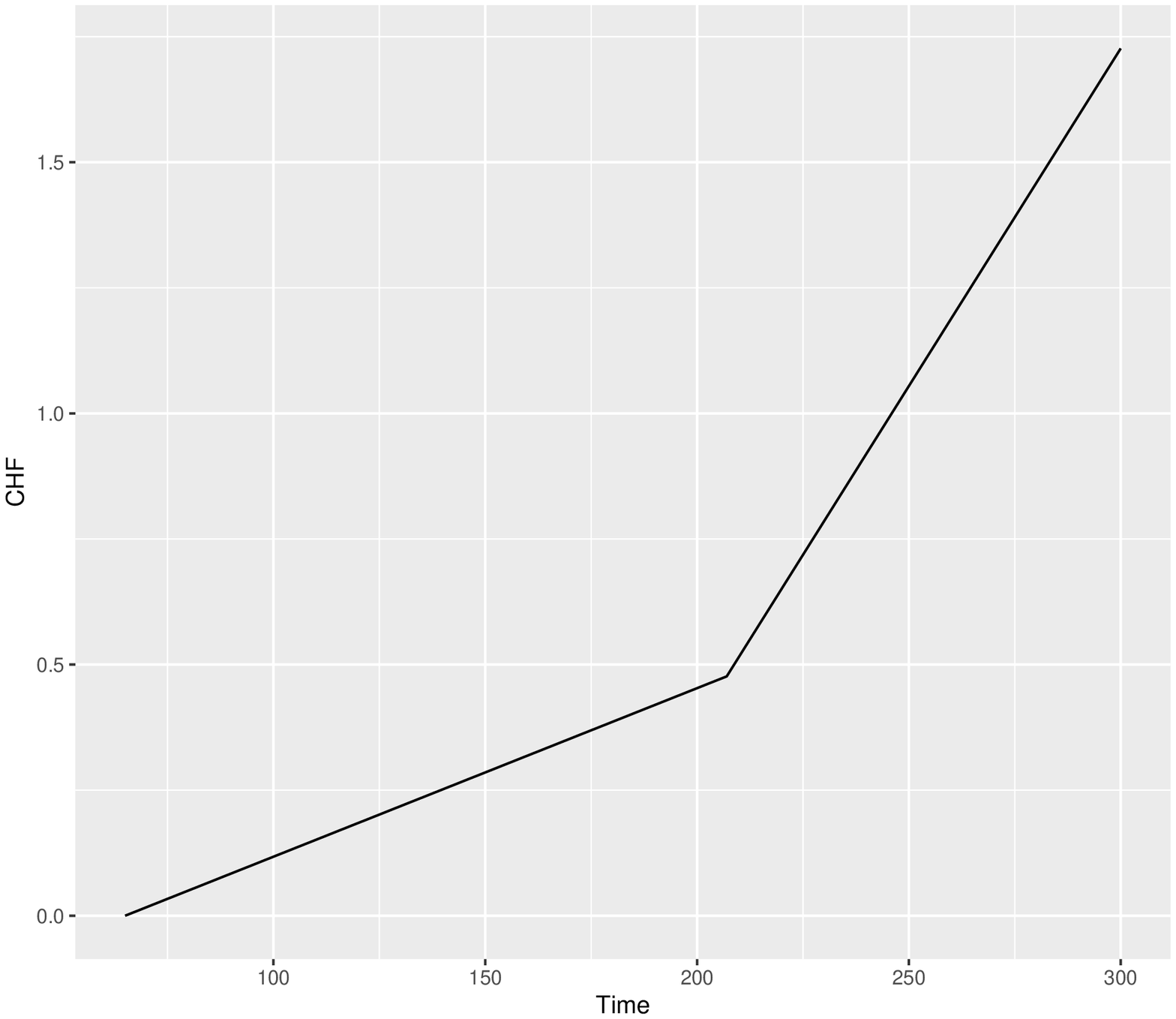}  
		\caption{Plot of CHF for $Y^{(0)}$}
		\label{fig:sub-first3}
	\end{subfigure}
	\begin{subfigure}{.5\textwidth}
		\centering
		\includegraphics[width=\linewidth]{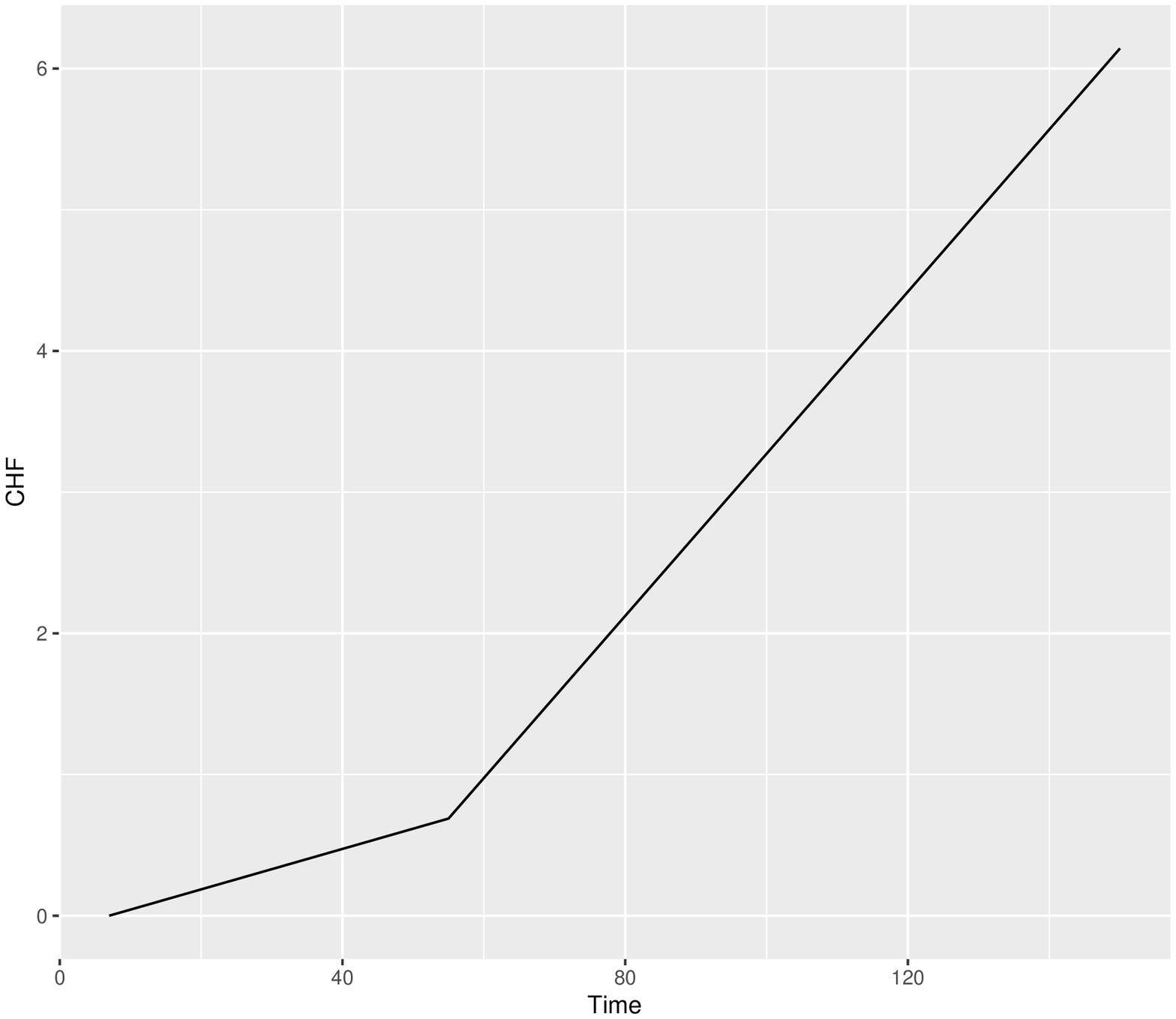}  
		\caption{Plot of CHF for $Y^{(1)}$}
		\label{fig:sub-second3}
	\end{subfigure}
	\caption{Plots of CHFs}
	\label{fig:fig3}
\end{figure}

\begin{table}[h!]
 \caption{Time to failure (in days) data set for two motors in a load-sharing configuration}
 \vspace{0.2cm}
 \label{tab:realdata}
 \centering
 \begin{tabular}{cccc}
    \hline
    System & Time to failure of motor A & Time to failure of motor B & Event description\\
   \hline
    1 & 102 & 65 & B failed first\\
    2 & 84 & 148 & A failed first\\
    3 & 88 & 202 & A failed first\\
    4 & 156 & 121 & B failed first\\
    5 & 148 & 123 & B failed first\\
    6 & 139 & 150 & A failed first\\
    7 & 245 & 156 & B failed first\\
    8 & 235 & 172 & B failed first\\
    9 & 220 & 192 & B failed first\\
    10 & 207 & 214 & A failed first\\
    11 & 250 & 212 & B failed first\\
    12 & 212 & 220 & A failed first\\
    13 & 213 & 265 & A failed first\\
    14 & 220 & 275 & A failed first\\
    15 & 243 & 300 & A failed first\\
    16 & 300 & 248 & B failed first\\
    17 & 257 & 330 & A failed first\\
    18 & 263 & 350 & A failed first\\
    \hline
 \end{tabular}
\end{table}

\begin{table}[h]
   \footnotesize
   \centering
   \caption{Point and interval estimates of parameters of the PLA-based model when applied to the two-motor load-sharing data}
   \label{tab:dataresult}
   \begin{tabular}{*{6}{c}}
      \toprule
      Parameter & MLE & Std.~Error & Asymptotic & Percentile bootstrap & Bootstrap \\
      \midrule
      $\gamma_1$ & 4.2712  & 1.1901 & (1.9386, 6.6038) & (3.0754, 8.0279) & (0.8456, 5.8172)\\
      $b_1$ & 0.0034 & 0.0008 & (0.0019, 0.0048) & (0.0021, 0.0062) & (0.0008,  0.0052) \\
      $b_2$ & 0.0134 &  0.0039 & (0.0056, 0.0212) & (0.0061, 0.0209) & (0.0083, 0.0232)\\
      \bottomrule
   \end{tabular}
\end{table}
 
\begin{table}[h!]
 \caption{Mean residual time and reliability in mission time}
 \vspace{0.2cm}
 \label{tab:RMT_MRT}
 \centering
\begin{tabular}{ccc}
\hline
$t_0$ & MRT$_{t_0}$  & RMT$_{t_0}$\\

\hline
102.00 & 124.223 & 0.963\\
167.50 & 88.678 & 0.706\\
227.50 & 60.646 & 0.466\\
272.50 & 42.794 & 0.271\\
350.00 & 36.919 & 0.044\\
\hline
     \end{tabular}
\end{table}

Sutar and Naik-Nimbalkar~\cite{SN2014} observed that the load-sharing phenomenon existed for the systems considered in this dataset. Asha et al.~\cite{ARR2018} assumed Weibull lifetimes for the components. From the Weibull Q-Q plots for the lifetimes of motor A and B reported in Asha et al.~\cite{ARR2018}, it was observed that although the Weibull model assumption for the lifetimes of motor B was reasonable, the lifetimes of motor A did not follow a Weibull distribution. This motivated us to consider the PLA-based modelling approach for the lifetimes of the load-sharing systems in this case.      

The dataset is reproduced in Table~\ref{tab:realdata} for ready reference of the readers.
The average and standard deviation of first component failure times are 178.61 and 62.75,
respectively, while those of the lifetime between first and second component failures are
49.72 and 29.45, respectively. We consider three cut points for the PLA-based model (i.e.,
$N=2$). The estimates of the model parameters are reported in Table~\ref{tab:dataresult}.
The Q-Q plots for $Y^{(0)}$ and $Y^{(1)}$ are given in Figures~\ref{fig:sub-qq0} and \ref{fig:sub-qq1}, respectively. The plots of the estimated SF and CHF are given in Figures~\ref{fig:fig2} and \ref{fig:fig3}, respectively. These figures indicate that the PLA-based model fits the data quite adequately.

A Kolmogorov-Smirnov type test has been performed to test the following hypotheses:
\begin{equation*}
\begin{array}{c}
H_0:\text{ True model is specified by Eqs.~\eqref{eq:model1} and \eqref{eq:model2}}\\
\text{against}\\
H_1: \text{ True model is not specified by Eqs.~\eqref{eq:model1} and \eqref{eq:model2}}
\end{array}
\end{equation*}
based on the test statistics
\begin{align*}
   T_n= \max_{1\le i\le n} \left\vert \widehat{G}^{(0)}\left( Y^{(0)}_{i:n} \right) -
   \frac{i}{n} \right\vert + \max_{1\le i\le n} \left\vert \widehat{G}^{(1)}\left(
   Y^{(1)}_{i:n} \right) - \frac{i}{n} \right\vert,
\end{align*}
where $\widehat{G}^{(j)}(\cdot)$ is the estimated cumulative distribution function
corresponding to PLA-based model, and $Y^{(j)}_{i:n}$ is the $i$-th order statistics
corresponding to $Y^{(j)}_i$, $j=0,\,1$,  $i=1,\,2,\,\ldots,\,n$. The observed value of the test statistics $T_n$ is found to be 0.414 based on this data. The Monte Carlo estimate of the corresponding $p$-value is 0.71. Therefore, the null hypothesis cannot be rejected at significance level 0.05, and we conclude that it is quite reasonable to use the PLA-based model for this data.  

It may also be noted here that for this data, the value of the Akaike's information criterion (AIC) for the model considered by Asha et al.~\cite{ARR2018} is 480.50, and that for the best model considered by Franco et~al.~\cite{FVK2020} is 409.65. In contrast, the AIC value for the PLA-based model turns out to be 369.34, implying that the PLA-based model is more suitable for the two-motor load-sharing systems data considered here.

For the PLA-based model, the estimated value of $\gamma_1$ is 4.2712, which empirically implies that the load-sharing model is quite appropriate in this case. The same comment can also be made from the plots, by noting that the plot of the SF of the distribution of time between first and second failure component times diminishes to zero more quickly compared to that of first component failure times in Figure~\ref{fig:fig2}.

The reliability characteristics of the two-motor load-sharing systems are also estimated by using the expressions and techniques described in Section \ref{sec:RC}. The MTTF is calculated to be 221.36 days. Monte Carlo estimates of the MRT and RMT are calculated at different sample percentile points of the system failure times and are presented in Table \ref{tab:RMT_MRT}.

\section{\sc Simulation Study} \label{sec:sim}
The accuracy of the proposed PLA-based model in fitting data from load-sharing systems is
of utmost importance as the subsequent estimation of reliability characteristics depends
on the PLA-based model. In this section, we present results of a Monte Carlo simulation
study that examines the performance of the proposed PLA-based model in two directions.
First, based on samples generated from a parent process with piecewise linear CHF, we
assess the performance of the proposed estimation method that is presented in Section
\ref{sec:Inf}. Then, the efficacy of the PLA-based model in fitting data generated from a
parent process represented by some parametric models is also assessed. The simulations are
carried out by using \texttt{R} software. For the simulations, we consider two-component
load-sharing systems.

\subsection{\sc Assessing performance of the estimation method}
To assess the performance of the estimation methods, we consider an underlying cumulative
hazard that is made up of two linear pieces. To this effect, we generate samples from the
model specified by Eqs.\eqref{eq:model1} and \eqref{eq:model2} with $J=2$ and $N=2$. The
true parameter values are taken to be $b_1=0.01,\, 0.05$; $b_2 = 0.1,\,0.5$; $\gamma_1=5$;
$\tau_1^{(0)}=\frac{\ln 2}{2b_1}$; $\tau_1^{(1)}=\frac{\ln 2}{\gamma_1 b_1}$. The
estimation is performed based on samples of size $n= 100$ and 200. The average estimates
(AE), mean square errors (MSE), variance (VAR) of the MLEs based on 5000 Monte Carlo
replications are reported in Tables~\ref{tab:sim1}, \ref{tab:sim2}, and \ref{tab:sim3}.
The coverage percentage (CP) and average lengths (AL) of 95\% confidence intervals are
also reported in the same tables. 

From the Tables \ref{tab:sim1}, \ref{tab:sim2} and \ref{tab:sim3}, we observe that the
average estimates of $\gamma_1,~ b_1$ and $b_2$ are very close to the true values, and the
MSEs as well as VARs are quite small as desired. It is also noticed that the performance
of all the constructed confidence intervals is satisfactory. These results demonstrate
that the proposed inferential techniques can accurately estimate the parameters of the
PLA-based model.  

\begin{table}[h]
	\footnotesize
	\centering
	\caption{Performance measures for estimates of $\gamma_1$}
	\begin{tabular}{c|ccccccccccc}
		\toprule
		\multirow{2}{1em}{$n$}&\multirow{2}{1em}{$b_1$}&\multirow{2}{1em}{$b_2$}&\multirow{2}{2em}{AE}&\multirow{2}{2.5em}{MSE}&\multirow{2}{2.5em}{VAR}&\multicolumn{2}{c}{Asymptotic}& \multicolumn{2}{c}{Percentile bootstrap}&\multicolumn{2}{c}{Bootstrap}\\
		\cmidrule(lr){7-8}	\cmidrule(lr){9-10} \cmidrule(lr){11-12}
       &&&&&&CP & AL & CP & AL & CP & AL\\
		\midrule

		&\multirow{2}{1em}{0.01}& 0.1 & 5.0231 & 0.3388811 & 0.3384155 & 94.38 & 2.2566 & 99.94 & 2.2012 & 83.58 & 2.2156 \\ 
		& 	& 0.5 & 5.0178 & 0.2880872 & 0.2878297 & 95.84 & 2.2509 & 99.94 & 2.1278 & 86.68 & 2.1993 \\
		\cmidrule(lr){2-12}
		100	& \multirow{2}{1em}{0.05}& 0.1 & 5.0209 & 0.3556721 & 0.3553026 & 93.98 & 2.2711 & 98.52 & 2.3093 & 88.60 & 2.3226 \\
		& 	& 0.5 & 5.0231 & 0.3388811 & 0.3384155 & 94.38 & 2.2566 & 99.94 & 2.2012 & 83.58 & 2.2156 \\
		\midrule
		&\multirow{2}{1em}{0.01}& 0.1 & 5.0144 & 0.1472563 & 0.1470773 & 96.20 & 1.5963 & 99.90 & 1.5000 & 85.06 & 1.5093 \\
		& 	& 0.5 & 5.0127 & 0.1373738 & 0.1372388 &  96.64 & 1.5949 & 99.86 & 1.4395 & 84.42 & 1.4476 \\
		\cmidrule(lr){2-12}
		200	&	\multirow{2}{1em}{0.05}& 0.1 & 5.0145 & 0.1698002 & 0.1696241 & 94.84 & 1.6037 & 98.56 & 1.6165 & 89.56 & 1.6246 \\

		& 	& 0.5 &  5.0144 & 0.1472563 & 0.1470773 & 96.20 & 1.5963 & 99.90 & 1.5000 & 85.06 & 1.5093 \\
		\bottomrule
	\end{tabular}
	\label{tab:sim1}
\end{table}

\begin{table}[h]
	\footnotesize
	\centering
	\caption{Performance measures for estimates of $b_1$}
	\begin{tabular}{c|ccccccccccc}
		\toprule
		\multirow{2}{1em}{$n$}&\multirow{2}{1em}{$b_1$}&\multirow{2}{1em}{$b_2$}&\multirow{2}{2em}{AE}&\multirow{2}{2.5em}{MSE}&\multirow{2}{2.5em}{VAR}&\multicolumn{2}{c}{Asymptotic}& \multicolumn{2}{c}{Percentile bootstrap}&\multicolumn{2}{c}{Bootstrap}\\
		\cmidrule(lr){7-8}	\cmidrule(lr){9-10} \cmidrule(lr){11-12}
       &&&&&&CP & AL & CP & AL & CP & AL\\
		\midrule

		&\multirow{2}{1em}{0.01}& 0.1 & 0.0108 & 0.0000022 & 0.0000016 &  87.24 & 0.0041 & 81.66 & 0.0052 & 92.88 & 0.0052 \\
		& 	& 0.5 & 0.0110 & 0.0000025 & 0.0000016 & 84.56 & 0.0042 & 68.70 & 0.0053 & 93.96 & 0.0054 \\
		\cmidrule(lr){2-12}

		100	& \multirow{2}{1em}{0.05}& 0.1 & 0.0513 & 0.0000389 & 0.0000371 & 90.10 & 0.0201 & 95.96 & 0.0245 & 93.70 & 0.0246 \\
		& 	& 0.5 & 0.0528 & 0.0000457 & 0.0000376 & 89.18 & 0.0203 & 89.88 & 0.0252 & 92.70 & 0.0253 \\
		\midrule
		&\multirow{2}{1em}{0.01}& 0.1 & 0.0105 & 0.0000010 & 0.0000008 & 86.42 & 0.0029 & 81.74 & 0.0035 & 93.48 & 0.0036 \\
		& 	& 0.5 & 0.0106 & 0.0000011 & 0.0000008 &  84.74 & 0.0029 & 74.08 & 0.0036 & 94.56 & 0.0036 \\
		\cmidrule(lr){2-12}

		200	&	\multirow{2}{1em}{0.05}& 0.1 & 0.0508 & 0.0000186 & 0.0000180 & 90.06 & 0.0140 & 95.14 & 0.0169 & 93.08 & 0.0169 \\
		& 	& 0.5 & 0.0525 & 0.0000255 & 0.0000193 &  86.42 & 0.0143 & 81.74 & 0.0177 & 93.48 & 0.0178 \\
		\bottomrule
	\end{tabular}
	\label{tab:sim2}
\end{table}

\begin{table}[h]
	\footnotesize
	\centering
	\caption{Performance measures for estimates of $b_2$}
	\begin{tabular}{c|ccccccccccc}
		\toprule
		\multirow{2}{1em}{$n$}&\multirow{2}{1em}{$b_1$}&\multirow{2}{1em}{$b_2$}&\multirow{2}{2em}{AE}&\multirow{2}{2.5em}{MSE}&\multirow{2}{2.5em}{VAR}&\multicolumn{2}{c}{Asymptotic}& \multicolumn{2}{c}{Percentile bootstrap}&\multicolumn{2}{c}{Bootstrap}\\
		\cmidrule(lr){7-8}	\cmidrule(lr){9-10} \cmidrule(lr){11-12}
       &&&&&&CP & AL & CP & AL & CP & AL\\
		\midrule
		&\multirow{2}{1em}{0.01}& 0.1 & 0.1006 & 0.0001691 & 0.0001688 & 92.70 & 0.0464 & 96.60 & 0.0534 & 94.00 & 0.0530 \\
		& 	& 0.5 & 0.5067 & 0.0038339 & 0.0037895 &  94.52 & 0.2344 & 97.12 & 0.2675 & 95.12 & 0.2676 \\
		\cmidrule(lr){2-12}
		100& \multirow{2}{1em}{0.05}& 0.1 & 0.1030 & 0.0001794 & 0.0001705 & 93.76 & 0.0472 & 95.46 & 0.0529 & 94.70 & 0.0532 \\
		& 	& 0.5 &  0.5028 & 0.0042337 & 0.0042268 &  92.68 & 0.2315 & 96.34 & 0.2650 & 94.00 & 0.2633 \\
		\midrule
		&\multirow{2}{1em}{0.01}& 0.1 & 0.1002 & 0.0000725 & 0.0000725 & 94.22 & 0.0325 & 96.72 & 0.0343 & 93.76 & 0.0345 \\  
		& 	& 0.5 &  0.5030 & 0.0017348 & 0.0017261 & 95.06 & 0.1632 & 96.52 & 0.1665 & 93.60 & 0.1674\\
		\cmidrule(lr){2-12}
		200	&	\multirow{2}{1em}{0.05}& 0.1 & 0.1011 & 0.0000780 & 0.0000768 & 93.84 & 0.0326 & 96.10 & 0.0351 & 94.08 & 0.0352 \\ 

		& 	& 0.5 & 0.5010 & 0.0018134 & 0.0018128 & 94.22 & 0.1623 & 96.72 & 0.1717 & 93.76 & 0.1725 \\
		\bottomrule
	\end{tabular}
	\label{tab:sim3}
\end{table}

\begin{table}[h]
	\centering
	\caption{AIE based on SF and CHF for Weibull distribution with $k=3$, $\beta=1$.}
	\begin{tabular}{c|ccccc}
		\toprule
		$n$ & $\alpha$ & $AIE^{(0)}_{SF}$ & $AIE^{(1)}_{SF}$ & $AIE^{(0)}_{CHF}$ & $AIE^{(1)}_{CHF}$\\ 
		\midrule
		\multirow{2}{1em}{50}& 1.0 & 0.0379 & 0.0291 & 0.1503 & 0.2981 \\
		& 1.5 & 0.0436 & 0.0434 & 0.1329 & 0.2633 \\
		\midrule
		\multirow{2}{1.5em}{100}& 1.0 & 0.0266 & 0.0183 & 0.1282 & 0.2541 \\
		& 1.5 & 0.0326 & 0.0301 & 0.1231 & 0.2440 \\
		\bottomrule
	\end{tabular}
	\label{tab:rise_weibull}
\end{table}

\begin{table}[h]
	\centering
	\caption{AIE of the survival and cumulative hazard function of quadratic distribution for $\kappa_1=0.5,~\tilde\kappa_1=2\kappa_1=1,~ \tilde\kappa_2>2\kappa_2$.}
   \begin{tabular}{c|c@{\hspace{7mm}}ccccc}
		\toprule
		$n$ & $\kappa_2$ & $\tilde\kappa_2$ & $AIE_{SF}^{(0)}$ & $AIE_{SF}^{(1)}$ & $AIE_{CHF}^{(0)}$ & $AIE_{CHF}^{(1)}$\\
		\midrule
		\multirow{4}{1em}{50}& \multirow{2}{1em}{0.50} & 1.50 & 0.0380 & 0.0368 & 0.1262 & 0.2536 \\
		& & 2.00 & 0.0380 & 0.0389 & 0.1261 & 0.2555\\
		\cmidrule(lr){2-7}
		& \multirow{2}{1em}{0.70} & 1.50 & 0.0389 & 0.0363 & 0.1261 & 0.2524 \\
		& & 2.00 & 0.0388 & 0.0383 & 0.1258 & 0.2539\\
		\midrule
		\multirow{4}{1.5em}{100}& \multirow{2}{1em}{0.50} & 1.50 & 0.0289 & 0.0262 & 0.1185 & 0.2506 \\
		& & 2.00 & 0.0289 & 0.0281 & 0.1178 & 0.2575\\
		\cmidrule(lr){2-7}
		& \multirow{2}{1em}{0.70} & 1.50 & 0.0301 & 0.0257 & 0.1217 & 0.2465 \\
		& & 2.00 & 0.0299 & 0.0274 & 0.1206 & 0.2528 \\
		\bottomrule
	\end{tabular}
	\label{tab:rise_quad_beta}
\end{table}

\begin{table}[h]
	\centering
	\caption{AIE of the survival and cumulative hazard function of quadratic distribution for $\tilde\kappa_1>2\kappa_1,~ \kappa_2=0.5,~\tilde\kappa_2=2\kappa_2=1$.}
   \begin{tabular}{c|c@{\hspace{7mm}}ccccc}
		\toprule
		$n$ & $\kappa_1$ & $\tilde\kappa_1$ & $AIE_{SF}^{(0)}$ & $AIE_{SF}^{(1)}$ & $AIE_{CHF}^{(0)}$ & $AIE_{CHF}^{(1)}$\\
		\midrule
 
		\multirow{4}{1em}{50}& \multirow{2}{1em}{0.50} & 1.50 & 0.0388 & 0.0313 & 0.1283 & 0.2570 \\
		
		& & 2.00 & 0.0397 & 0.0307 & 0.1309 & 0.2672 \\
		
		\cmidrule(lr){2-7}
 
		& \multirow{2}{1em}{0.70} & 1.50 & 0.0372 & 0.0314 & 0.1284 & 0.2567 \\
		
		& & 2.00 & 0.0377 & 0.0304 & 0.1301 & 0.2644 \\
		
		\midrule
 
		\multirow{4}{1.5em}{100}& \multirow{2}{1em}{0.50} & 1.50 & 0.0306 & 0.0210 & 0.1265 & 0.2290 \\
		& & 2.00 & 0.0319 & 0.0206 & 0.1325 & 0.2285 \\
		\cmidrule(lr){2-7}
 
		& \multirow{2}{1em}{0.70} & 1.50 & 0.0278 & 0.0210 & 0.1184 & 0.2331 \\
		& & 2.00 & 0.0285 & 0.0198 & 0.1227 & 0.2271 \\
		\bottomrule
	\end{tabular}
	\label{tab:rise_quad_alpha}
\end{table}

\subsection{\sc Assessing efficacy of the PLA-based model in fitting data from other models}
Now, we examine the robustness of the PLA-based model in the following manner. We generate
load-sharing data from parametric models, and then fit the PLA-based model to the data.
The model fit is then assessed with respect to an integrated measure that is suitably
defined to reflect the quality of approximation provided by the PLA-based model. The
measure, which we call the Absolute Integrated Error (AIE), is as follows. For $j=0,\,1$,
let $S^{(j)}_{TGP}(\cdot)$ and $H^{(j)}_{TGP}(\cdot)$ denote the SF and CHF of the
lifetimes between $j$-th and $(j+1)$-st failures. Also, assume that the estimated SF and
CHF based on PLA-based model are denoted by $\widehat{S}^{(j)}_{PLA}(\cdot)$ and
$\widehat{H}^{(j)}_{PLA}(\cdot)$, respectively. Then the AIE, based on the SF and CHF,
respectively, are defined as
\begin{align*}
   AIE_{SF}^{(j)} &= \frac{1}{R}\sum_{k=1}^{R}
   \frac{1}{y^{(j)}_{\max} - y^{(j)}_{\min}} \int_{y^{(j)}_{\min}}^{y^{(j)}_{\max}}
   \left\vert S_{TGP}^{(j)}(t)-\widehat{S}_{PCA}^{(j)}(t)\right\vert dt, \\[2mm]
   AIE_{CHF}^{(j)} &= \frac{1}{R}\sum_{k=1}^{R}
   \frac{1}{y^{(j)}_{\max} - y^{(j)}_{\min}} \int_{y^{(j)}_{\min}}^{y^{(j)}_{\max}}
   \left\vert H_{TGP}^{(j)}(t)-\widehat{H}_{PCA}^{(j)}(t)\right\vert dt,
\end{align*}
where $y^{(j)}_{\min} = \min\left\{ y^{(j)}_1,\, y^{(j)}_2,\, \ldots,\,
y^{(j)}_n \right\}$, $y^{(j)}_{\max} = \max\left\{ y^{(j)}_1,\, y^{(j)}_2,\, \ldots,\,
y^{(j)}_n \right\}$, $j=0,\,1$. 

For generating load-sharing data from parametric models, two scenarios are considered: \newline
(a) Case - 1: It is assumed that the lifetimes of each components of a two-component
load-sharing system are independent and identically distributed as Weibull distribution
with shape parameter $\alpha$ and scale parameter $\beta$ when both the components are
working. After the first failure, the lifetime of the surviving component is assumed to
follow a Weibull distribution with same shape parameter $\alpha$, but a different scale
parameter $k\beta$, where $k>2$ is to ensure the increase of load on the surviving
component. For $\beta=1$, $k=3$, we take $\alpha=1$ and 1.5.\\[-2mm]

\noindent (b) Case - 2: In the second scenario, the component lifetimes are assumed to be
independent and identically distributed according to a distribution with quadratic CHF
$\kappa_1 t+\kappa_2 t^2$ when both components are working. After the first failure, the
lifetime of the surviving component is assumed to follow a quadratic CHF with different
parameters $\tilde\kappa_1$ and $\tilde\kappa_2$. We take several values of the parameters
$\kappa_1,\,\kappa_2,\,\tilde\kappa_1,$ and $\tilde\kappa_2$ ensuring the fact that the
CHF increases after one component fails in the system. 

The numerical results are reported in Tables~\ref{tab:rise_weibull},
\ref{tab:rise_quad_beta}, and \ref{tab:rise_quad_alpha}. For all cases, it is observed
that the values of AIE based on SF and CHF are reasonably small, indicating that the
PLA-based model provides quite a satisfactory approximation to the data generated from
different parent populations.


\section{\sc Concluding Remarks} \label{sec:con}
In this article, a PLA-based model for the CHF is proposed for data from load-sharing systems and then important reliability characteristics such as quantile function, RMT, MTTF, and MRT of load-sharing systems are estimated under the proposed model. The principal advantages of the model are that it is data-driven, and does not use strong parametric assumptions for the underlying lifetime variable. Likelihood inference for the proposed model is discussed in detail. It is observed that for two-component load-sharing systems, it is possible to obtain explicit expressions for the MLEs of parameters of the PLA-based model. Construction of confidence intervals using the Fisher information matrix and bootstrap approaches are also discussed. Derivations of the important reliability characteristics are provided in this setting.  

A Monte Carlo simulation study is performed to examine (a) the performance of the methods of inference, and (b) the efficacy of the PLA-based model to fit load-sharing data in general. It is shown that the PLA-based model performs quite satisfactorily in both cases. Analysis of data pertaining to components lifetimes of a two-motor load-sharing system is provided as an illustration. It is illustrated that the PLA-based model is superior to the models that have been considered for this data in the literature of load-sharing systems. In summary, in this paper, an efficient PLA-based modelling framework using minimal assumptions for load-sharing systems is discussed, and estimates of important reliability characteristics for load-sharing systems in this setting are developed.

\section*{\sc Funding information}
\begin{itemize}
\item The research of Ayon Ganguly is supported by the Mathematical Research Impact
Centric Support (File no.~MTR/2017/000700) from the Science and
Engineering Research Board, Department of Science and Technology, Government of
India.
\item The research of Debanjan Mitra is supported by the Mathematical Research Impact
Centric Support (File no.~MTR/2021/000533) from the Science and
Engineering Research Board, Department of Science and Technology, Government of
India.
\end{itemize}

\section*{\sc Appendix A: Calculation of Fisher information matrix for two-component load sharing systems}
For calculating $\mathbb{I}(\boldsymbol \theta)$, the required expectations are $E\left(N_1^{(0)}\right),  E\left(N_2^{(0)}\right), E\left(N_1^{(1)}\right),  E\left(N_2^{(1)}\right),\\ E\left(\displaystyle \sum_{i \in I_1^{(1)}}Y_i^{(1)}\right)$ and $E\left(\displaystyle \sum_{i \in I_2^{(1)}}Y_i^{(1)}\right)$.\\
Note that 
$$N_k^{(0)} \sim Bin(n,p_k^{(0)}), \quad N_k^{(1)} \sim Bin(n, p_k^{(1)}),$$
with 
$$p_k^{(0)} = P\bigg(Y_i^{(0)} \in [\tau_{k-1}^{(0)}, \tau_k^{(0)})\bigg), \quad p_k^{(1)} = P\bigg(Y_i^{(1)} \in [\tau_{k-1}^{(1)}, \tau_k^{(1)})\bigg), \quad k=1, 2.$$ 
In case of a two-component load-sharing system, PDF of $Y_i^{(j)}$, $j=1,2$, is given by
   \begin{align*}
      g_{Y_i^{(j)}}(y)=(2-j)\lambda^{(j)}(y)e^{-(2-j)\int_{0}^{y}\lambda^{(j)}(u)du}.
   \end{align*}
Hence,
\begin{align*}
 p_1^{(0)} = \int_{0}^{\tau_1^{(0)}}g_{Y_i^{(0)}}(y)dy = 1-e^{-2b_1\tau_1^{(0)}}, \quad p_1^{(1)} = \int_{0}^{\tau_1^{(1)}}g_{Y_i^{(1)}}(y)dy = 1-e^{-\gamma_1 b_1\tau_1^{(1)}}.   
\end{align*}
Then, $p_2^{(0)} = 1-p_1^{(0)} = e^{-2b_1\tau_1^{(0)}}$ and $p_2^{(1)} = 1-p_1^{(1)} =
e^{-\gamma_1 b_1\tau_1^{(1)}}$. Therefore, $$E(N_1^{(0)}) = 1-e^{-2b_1\tau_1^{(0)}}, \quad
E(N_2^{(0)}) =e^{-2b_1\tau_1^{(0)}}, \quad E(N_1^{(1)}) = 1-e^{-\gamma_1 b_1\tau_1^{(1)}},
\quad E(N_2^{(1)}) = e^{-\gamma_1 b_1\tau_1^{(1)}}.$$
Now, 
$$ E\left(\sum_{i \in I_1^{(1)}}Y_i^{(1)}\right) = E\left(E\left(\sum_{i \in
I_1^{(1)}}Y_i^{(1)}|N_1^{(1)}=n_1^{(1)}\right)\right).$$
For $i\in I_1^{(1)}$, $Y_i^{(1)}$ follows a right truncated exponential distribution with
PDF $\frac{\gamma_1 b_1e^{-\gamma_1 b_1y}}{1-e^{-\gamma_1 b_1\tau_1^{(1)}}}$ for
$0<y<\tau_1^{(1)}$. Hence, for $i\in I_1^{(1)}$, 
$$ E\left(Y_i^{(1)}\right) = \int_{0}^{\tau_1^{(1)}}y\frac{\gamma_1 b_1e^{-\gamma_1 b_1y}}{1-e^{-\gamma_1 b_1\tau_1^{(1)}}}dy = \frac{1}{\gamma_1 b_1}\left[\frac{1-(1+\gamma_1 b_1 \tau_1^{(1)})e^{-\gamma_1 b_1\tau_1^{(1)}}}{1-e^{-\gamma_1 b_1\tau_1^{(1)}}}\right]. $$
Therefore, 
   \begin{align*}
      E\left(\sum_{i \in I_1^{(1)}}Y_i^{(1)}\right)&=\frac{1}{\gamma_1 b_1}\left[\frac{1-(1+\gamma_1 b_1 \tau_1^{(1)})e^{-\gamma_1 b_1\tau_1^{(1)}}}{1-e^{-\gamma_1 b_1\tau_1^{(1)}}}\right] E(N_1^{(1)})\\&=\frac{1}{\gamma_1 b_1}\left[\frac{1-(1+\gamma_1 b_1 \tau_1^{(1)})e^{-\gamma_1 b_1\tau_1^{(1)}}}{1-e^{-\gamma_1 b_1\tau_1^{(1)}}}\right] \left(1-e^{-\gamma_1 b_1\tau_1^{(1)}}\right)\\&=\frac{1}{\gamma_1 b_1}\left[{1-(1+\gamma_1 b_1 \tau_1^{(1)})e^{-\gamma_1 b_1\tau_1^{(1)}}}\right].
   \end{align*}
Similarly,
$$ E\left(\sum_{i \in I_2^{(1)}}Y_i^{(1)}\right) = E\left(E\left(\sum_{i \in I_2^{(1)}}Y_i^{(1)}|N_2^{(1)}=n_2^{(1)}\right)\right).$$
For $i\in I_2^{(1)}$, $Y_i^{(1)}$ follows a left truncated exponential distribution with PDF 
$\frac{\gamma_1 b_2e^{-\gamma_1 b_2y}}{e^{-\gamma_1 b_2\tau_1^{(1)}}}$ for
$y>\tau_1^{(1)}$. Hence, 
$$ E\left(Y_i^{(1)}\right) = \int_{\tau_1^{(1)}}^{\infty}y\frac{\gamma_1 b_2e^{-\gamma_1 b_2y}}{e^{-\gamma_1 b_2\tau_1^{(1)}}}dy = \frac{1}{\gamma_1 b_2}+\tau_1^{(1)}.$$
Therefore,
$$ E\left(\sum_{i \in I_2^{(1)}}Y_i^{(1)}\right) = \left(\frac{1}{\gamma_1 b_2}+\tau_1^{(1)}\right) E(N_2^{(1)}) = \left(\frac{1}{\gamma_1 b_2}+\tau_1'\right)e^{-\gamma_1 b_1\tau_1^{(1)}}. $$

\section*{\sc Appendix B: Calculations of some important reliability characteristics}
\textbf{Derivation of the quantile function:}

Denote $p=G^{(j)}(y)$ for $y\in \left[\tau_{k-1}^{(j)}, \tau_k^{(j)}\right)$; then,
$y=\eta(p)$ for $p\in \left[G^{(j)}(\tau_{k-1}^{(j)}), G^{(j)}(\tau_{k}^{(j)})\right)$, $k=1, 2, \ldots, N.$
Now,
\begin{align*}
    p &= 1-e^{-(J-j)\gamma_j\left[\sum_{\ell=1}^{k-1}b_\ell\left(\tau_\ell^{(j)}-\tau_{\ell-1}^{(j)}\right)+b_k\left(y-\tau_{k-1}^{(j)}\right)\right]} \\
    \implies& b_k\left(y-\tau_{k-1}^{(j)}\right) = -\frac{\log(1-p)}{(J-j)\gamma_j}-\sum_{\ell=1}^{k-1}b_\ell\left(\tau_\ell^{(j)}-\tau_{\ell-1}^{(j)}\right) \\
    \implies& y =
    \tau_{k-1}^{(j)}-\frac{\log(1-p)}{(J-j)\gamma_jb_k}-\frac{1}{b_k}\sum_{\ell=1}^{k-1}b_\ell\left(\tau_\ell^{(j)}-\tau_{\ell-1}^{(j)}\right),
    \text{ if }  p\in \left[G^{(j)}(\tau_{k-1}^{(j)}), G^{(j)}(\tau_{k}^{(j)})\right),\\
            & \hspace{9cm} k=1, 2, \ldots, N.
\end{align*} 
If $y\in \left[\tau_{N}^{(j)}, \infty \right)$, then $y=\eta(p)$ for $p\in \left[G^{(j)}(\tau_{N}^{(j)}), 1\right)$.\\
Therefore,
\begin{align*}
p &= 1-e^{-(J-j)\gamma_j\left[\sum_{\ell=1}^{N-1}b_\ell\left(\tau_\ell^{(j)}-\tau_{\ell-1}^{(j)}\right)+b_N\left(y-\tau_{N-1}^{(j)}\right)\right]} \\
 \implies& y = \tau_{N-1}^{(j)}-\frac{\log(1-p)}{(J-j)\gamma_jb_N}-\frac{1}{b_N}\sum_{\ell=1}^{N-1}b_\ell\left(\tau_\ell^{(j)}-\tau_{\ell-1}^{(j)}\right), \text{ if } p\in \left[G^{(j)}(\tau_{N}^{(j)}), 1\right) .
\end{align*}

\noindent \textbf{Derivation of MTTF:}

MTTF of the system lifetime $T$ is given by $E(T)=E\left(\displaystyle\sum_{j=0}^{J-1}Y^{(j)}\right)=\displaystyle\sum_{j=0}^{J-1}E(Y^{(j)})$, where
\begin{equation}
E(Y^{(j)}) = \int_0^\infty P(Y^{(j)}>y)dy = \int_0^{\tau_{N-1}^{(j)}}e^{-(J-j)\Lambda^{(j)}(y)}dy+\int_{\tau_{N-1}^{(j)}}^{\infty}e^{-(J-j)\Lambda^{(j)}(y)}dy = I_1+I_2 \; (\textrm{say}). \nonumber
\end{equation}
Here,
\begin{align*}
  I_1&= \int_0^{\tau_{N-1}^{(j)}}e^{-(J-j)\gamma_j\sum_{k=1}^{N}\left[\sum_{\ell=1}^{k-1}b_\ell\left(\tau_\ell^{(j)}-\tau_{\ell-1}^{(j)}\right)+b_k\left(y-\tau_{k-1}^{(j)}\right)\right]\boldsymbol{1}_{[\tau^{(0)}_{k-1},\,
   \tau^{(0)}_{k})} \left( y \right)}dy\\
   &=\sum_{s=1}^{N-1}\int_{\tau_{s-1}^{(j)}}^{\tau_{s}^{(j)}}e^{-(J-j)\gamma_j\left[\sum_{\ell=1}^{s-1}b_\ell\left(\tau_\ell^{(j)}-\tau_{\ell-1}^{(j)}\right)+b_s\left(y-\tau_{s-1}^{(j)}\right)\right]}dy\\
   &=\sum_{s=1}^{N-1}\left\{e^{-(J-j)\gamma_j\sum_{\ell=1}^{s-1}b_\ell\left(\tau_\ell^{(j)}-\tau_{\ell-1}^{(j)}\right)}\int_{\tau_{s-1}^{(j)}}^{\tau_{s}^{(j)}}e^{-(J-j)\gamma_jb_s\left(y-\tau_{s-1}^{(j)}\right)}dy\right\}\\
   &=\sum_{s=1}^{N-1}\left\{e^{-(J-j)\gamma_j\sum_{\ell=1}^{s-1}b_\ell\left(\tau_\ell^{(j)}-\tau_{\ell-1}^{(j)}\right)}\left[\frac{1-e^{-(J-j)\gamma_jb_s\left(\tau_s^{(j)}-\tau_{s-1}^{(j)}\right)}}{(J-j)\gamma_j b_s}\right]\right\}\\
   &=\sum_{s=1}^{N-1}\left\{\frac{e^{-(J-j)\gamma_j\sum_{\ell=1}^{s-1}b_\ell\left(\tau_\ell^{(j)}-\tau_{\ell-1}^{(j)}\right)}-e^{-(J-j)\gamma_j\sum_{\ell=1}^{s}b_\ell\left(\tau_\ell^{(j)}-\tau_{\ell-1}^{(j)}\right)}}{(J-j)\gamma_j b_s}\right\}
\end{align*}
and
\begin{align*}
  I_2&= \int_{\tau_{N-1}^{(j)}}^{\infty}e^{-(J-j)\gamma_j\sum_{k=1}^{N}\left[\sum_{\ell=1}^{k-1}b_\ell\left(\tau_\ell^{(j)}-\tau_{\ell-1}^{(j)}\right)+b_k\left(y-\tau_{k-1}^{(j)}\right)\right]\boldsymbol{1}_{[\tau^{(0)}_{k-1},\,
   \tau^{(0)}_{k})} \left( y \right)}dy\\
   &=\int_{\tau_{N-1}^{(j)}}^{\infty}e^{-(J-j)\gamma_j\left[\sum_{\ell=1}^{N-1}b_\ell\left(\tau_\ell^{(j)}-\tau_{\ell-1}^{(j)}\right)+b_N\left(y-\tau_{N-1}^{(j)}\right)\right]}dy\\
   &=e^{-(J-j)\gamma_j\sum_{\ell=1}^{N-1}b_\ell\left(\tau_\ell^{(j)}-\tau_{\ell-1}^{(j)}\right)}\int_{\tau_{N-1}^{(j)}}^{\infty}e^{-(J-j)\gamma_jb_N\left(y-\tau_{N-1}^{(j)}\right)}dy\\
   &=e^{-(J-j)\gamma_j\sum_{\ell=1}^{N-1}b_\ell\left(\tau_\ell^{(j)}-\tau_{\ell-1}^{(j)}\right)}\left[\frac{1}{(J-j)\gamma_j b_N}\right]\\
   &=\frac{e^{-(J-j)\gamma_j\sum_{\ell=1}^{N-1}b_\ell\left(\tau_\ell^{(j)}-\tau_{\ell-1}^{(j)}\right)}}{(J-j)\gamma_j b_N}.
\end{align*}
Therefore, 
\begin{align*}
  E(Y^{(j)})=\sum_{s=1}^{N}
  \left\{\frac{e^{-(J-j)\gamma_j\sum_{\ell=1}^{s-1}b_\ell\left(\tau_\ell^{(j)}-\tau_{\ell-1}^{(j)}\right)}-e^{-(J-j)\gamma_j\sum_{\ell=1}^{s}b_\ell\left(\tau_\ell^{(j)}-\tau_{\ell-1}^{(j)}\right)}}{(J-j)\gamma_j
  b_s}\right\}.
\end{align*}
From here, the results follows immediately. 

\noindent \textbf{Derivation of moment generating function of system lifetime:} 

Note that the system lifetime MGF of $T$ is $T=\displaystyle\sum_{j=0}^{J-1}Y^{(j)}$, where $Y^{(j)}$'s are independent for $j=0,1,\ldots, (J-1)$. Therefore, the MGF of $T$ is  
$\phi_T(t)=\displaystyle\prod_{j=0}^{J-1}\phi_{Y^{(j)}}(t)$. 
Now,
\begin{eqnarray}
& \phi_{Y^{(j)}}(t) & = E(e^{tY^{(j)}}) = \int_0^{\infty}e^{ty}g_{Y^{(j)}}(y)dy	\nonumber \\
&& = \int_0^{\tau_{N-1}^{(j)}}e^{ty}(J-j)\lambda^{(j)}(y)e^{-(J-j)\Lambda^{(j)}(y)}dy+\int_{\tau_{N-1}^{(j)}}^{\infty}e^{ty}(J-j)\lambda^{(j)}(y)e^{-(J-j)\Lambda^{(j)}(y)}dy	\nonumber \\&& = I_1+I_2 \text{ (say) }, \nonumber
\end{eqnarray} where $g_{Y^{(j)}}(y)=(J-j)\lambda^{(j)}(y)e^{-(J-j)\Lambda^{(j)}(y)}.$
For $t\in \mathbb{R}$,
\begin{align*}
I_1&= \int_0^{\tau_{N-1}^{(j)}}e^{ty}(J-j)\gamma_j
\sum_{k=1}^{N}b_k\boldsymbol{1}_{[\tau^{(j)}_{k-1},\,
	\tau^{(j)}_{k})} \left( y \right)
   e^{-(J-j)\gamma_j\sum_{k=1}^{N}\left[\sum_{\ell=1}^{k-1}b_\ell\left(\tau_\ell^{(j)}-\tau_{\ell-1}^{(j)}\right)+b_k\left(y-\tau_{k-1}^{(j)}\right)\right]}dy\\
&=\sum_{s=1}^{N-1}(J-j)b_s\gamma_j
\int_{\tau_{s-1}^{(j)}}^{\tau_{s}^{(j)}}e^{-\left\{(J-j)\gamma_j\left[\sum_{\ell=1}^{s-1}b_\ell\left(\tau_\ell^{(j)}-\tau_{\ell-1}^{(j)}\right)+b_s\left(y-\tau_{s-1}^{(j)}\right)\right]-ty\right\}}dy\\
&=\sum_{s=1}^{N-1}\left\{(J-j)b_s\gamma_j e^{-(J-j)\gamma_j\sum_{\ell=1}^{s-1}b_\ell\left(\tau_\ell^{(j)}-\tau_{\ell-1}^{(j)}\right)}\int_{\tau_{s-1}^{(j)}}^{\tau_{s}^{(j)}}e^{-\left\{(J-j)\gamma_jb_s\left(y-\tau_{s-1}^{(j)}\right)-ty\right\}}dy\right\}\\
&=\sum_{s=1}^{N-1}\left\{(J-j)b_s\gamma_j
e^{-(J-j)\gamma_j\sum_{\ell=1}^{s-1}b_\ell\left(\tau_\ell^{(j)}-\tau_{\ell-1}^{(j)}\right)}\left[\frac{e^{t\tau_{s-1}^{(j)}}-e^{-\left\{(J-j)\gamma_jb_s\left(\tau_s^{(j)}-\tau_{s-1}^{(j)}\right)-t\tau_s^{(j)}\right\}}}{(J-j)\gamma_j b_s-t}\right]\right\}\\
&=\sum_{s=1}^{N-1}\frac{(J-j)b_s\gamma_j}{(J-j)
b_s\gamma_j-t}\left\{e^{-\left\{(J-j)\gamma_j\sum_{\ell=1}^{s-1}b_\ell\left(\tau_\ell^{(j)}-\tau_{\ell-1}^{(j)}\right)-t\tau_{s-1}^{(j)}\right\}}\right.\\
&\hspace{6cm}\left.-e^{-\left\{(J-j)\gamma_j\sum_{\ell=1}^{s}b_\ell\left(\tau_\ell^{(j)}-\tau_{\ell-1}^{(j)}\right)-t\tau_s^{(j)}\right\}} \right\}.
\end{align*}
For $t<(J-j)\gamma_j b_N$,
\begin{align*}
I_2&= \int_{\tau_{N-1}^{(j)}}^{\infty}e^{ty}(J-j)\gamma_j\sum_{k=1}^{N}b_k\boldsymbol{1}_{[\tau^{(j)}_{k-1},\,
	\tau^{(j)}_{k})} \left( y \right)
   e^{-(J-j)\gamma_j\sum_{k=1}^{N}\left[\sum_{\ell=1}^{k-1}b_\ell\left(\tau_\ell^{(j)}-\tau_{\ell-1}^{(j)}\right)+b_k\left(y-\tau_{k-1}^{(j)}\right)\right]}dy\\
&=(J-j)b_N\gamma_j
\int_{\tau_{N-1}^{(j)}}^{\infty}e^{ty-(J-j)\gamma_j\left[\sum_{\ell=1}^{N-1}b_\ell\left(\tau_\ell^{(j)}-\tau_{\ell-1}^{(j)}\right)+b_N\left(y-\tau_{N-1}^{(j)}\right)\right]}dy\\
&=(J-j)b_N\gamma_j
e^{-(J-j)\gamma_j\sum_{\ell=1}^{N-1}b_\ell\left(\tau_\ell^{(j)}-\tau_{\ell-1}^{(j)}\right)}\int_{\tau_{N-1}^{(j)}}^{\infty}e^{-\left\{(J-j)\gamma_jb_N\left(y-\tau_{N-1}^{(j)}\right)-ty\right\}}dy\\
&=(J-j)b_N\gamma_j
e^{-(J-j)\gamma_j\sum_{\ell=1}^{N-1}b_\ell\left(\tau_\ell^{(j)}-\tau_{\ell-1}^{(j)}\right)}\left[\frac{e^{t\tau_{N-1}^{(j)}}}{{(J-j)\gamma_j b_N-t}}\right]\\
&=(J-j)b_N\gamma_j\cdot \frac{e^{t\tau_{N-1}^{(j)}-(J-j)\gamma_j\sum_{\ell=1}^{N-1}b_\ell\left(\tau_\ell^{(j)}-\tau_{\ell-1}^{(j)}\right)}}{(J-j)b_N\gamma_j-t}.
\end{align*}
Therefore, for $t<(J-j)\gamma_j b_N$,
\begin{align*}
   \phi_{Y^{(j)}}(t)=&\sum_{s=1}^{N}\frac{(J-j)b_s\gamma_j}{(J-j)
 b_s\gamma_j-t}\left\{e^{-\left\{(J-j)\gamma_j\sum_{\ell=1}^{s-1}b_\ell\left(\tau_\ell^{(j)}-\tau_{\ell-1}^{(j)}\right)-t\tau_{s-1}^{(j)}\right\}}\right.\\
                     &\hspace{5cm}\left.-e^{-\left\{(J-j)\gamma_j\sum_{\ell=1}^{s}b_\ell\left(\tau_\ell^{(j)}-\tau_{\ell-1}^{(j)}\right)-t\tau_s^{(j)}\right\}}
                     \right\}.
\end{align*}
From here the result follows immediately. 
\end{document}